\documentclass[sigconf, screen]{acmart}
\AtBeginDocument{%
  }

\setcopyright{acmlicensed}
\copyrightyear{2018}
\acmYear{2018}
\acmDOI{XXXXXXX.XXXXXXX}
\acmConference[xxx]{xxx}{June 03--05,
  2018}{Woodstock, NY}
\acmISBN{978-1-4503-XXXX-X/2018/06}

\usepackage{booktabs}
\usepackage{array}
\usepackage{graphicx}
\usepackage{xspace}
\usepackage{amsmath}
\usepackage{tabu}
\usepackage{multirow}
\usepackage{enumitem}
\usepackage[table]{xcolor}
\usepackage{subcaption}

\usepackage{pifont}
\usepackage{tikz}
\definecolor{darkgreen}{RGB}{50,100,0}
\definecolor{darkred}{RGB}{200, 0, 0}
\newcommand{\cmark}{\textcolor{darkgreen}{\ding{51}}} 
\newcommand{\xmark}{\textcolor{darkred}{\ding{55}}} 

\usepackage[table, svgnames]{xcolor}
\definecolor{lightgreen}{rgb}{0.95, 1, 0.95}  
\definecolor{lightred}{rgb}{1, 0.95, 0.95}
\usepackage{tcolorbox}
\usepackage{tikz}

\newcommand{\eg}{\emph{e.g.,}\xspace}

\newcommand{\dname}{\textsc{M4DocBench}\xspace}
\newcommand{\mname}{\textsc{Doc-Researcher}\xspace}

\begin{document}

\title{Doc-Researcher: A Unified System for Multimodal Document Parsing and Deep Research}

\author{Kuicai Dong$^*$, Shurui Huang$^*$, Fangda Ye$^*$, Wei Han, Zhi Zhang, Dexun Li, Wenjun Li, Qu Yang, Gang Wang, Yichao Wang, Chen Zhang, Yong Liu}
\affiliation{%
  \institution{Huawei Technologies Co., Ltd.\\
  * denotes co-first authors;  \xspace correspond to \{dong.kuicai, liu.yong6\}@huawei.com
  \country{}}
}

\renewcommand{\shortauthors}{Dong et al.}

\begin{abstract}
    Deep Research systems have revolutionized how LLMs solve complex questions through iterative reasoning and evidence gathering. However, current systems remain fundamentally constrained to textual web data, overlooking the vast knowledge embedded in multimodal documents: scientific papers, technical reports, and financial documents where critical information exists in figures, tables, charts, and equations. Processing such documents demands sophisticated parsing to preserve visual semantics, intelligent chunking to maintain structural coherence, and adaptive retrieval across modalities, which are capabilities absent in existing systems. In response, we present \textbf{Doc-Researcher}, a unified system that bridges this gap through three integrated components: (i) \textbf{deep multimodal parsing} that preserves layout structure and visual semantics while creating multi-granular representations from chunk to document level, (ii) \textbf{systematic retrieval architecture} supporting text-only, vision-only, and hybrid paradigms with dynamic granularity selection, and (iii) \textbf{iterative multi-agent workflows} that decompose complex queries, progressively accumulate evidence, and synthesize comprehensive answers across documents and modalities.
    To enable rigorous evaluation, we introduce \textbf{M4DocBench}, the first benchmark for \textbf{M}ulti-modal, \textbf{M}ulti-hop, \textbf{M}ulti-document, and \textbf{M}ulti-turn deep research. Featuring 158 expert-annotated questions with complete evidence chains across 304 documents, M4DocBench tests capabilities that existing benchmarks cannot assess.
    Experiments demonstrate that Doc-Researcher achieves 50.6\% accuracy, 3.4× better than state-of-the-art baselines, validating that effective document research requires not just better retrieval, but fundamentally deep parsing that preserve multimodal integrity and support iterative research. Our work establishes a new paradigm for conducting deep research on multimodal document collections.

    
\end{abstract}



\received{20 February 2007}
\received[revised]{12 March 2009}
\received[accepted]{5 June 2009}

\maketitle

\section{Introduction}
\label{sec:introduction}

Recently, agentic retrieval-augmented generation (RAG) systems have significantly transformed the way LLMs retrieve, organize, and present information~\citep{singh2025agentic,xu2025comprehensivesurveydeepresearch}. Agentic RAG enhances traditional RAG paradigm by enabling systems to plan retrieval strategies, invoke external tools, adaptively refine queries, and validate context~\citep{li2025searcho1agenticsearchenhancedlarge}. These capabilities underpin recent Deep Research systems~\citep{huang2025deepresearchagentssystematic}, which support complex reasoning~\citep{li2025websailorv2bridgingchasmproprietary,li2025searcho1agenticsearchenhancedlarge,jin2025searchr,li2025webthinker0} and research workflows~\citep{li2025webweaverstructuringwebscaleevidence,du2025deepresearchbenchcomprehensivebenchmark}.
However, current deep research systems~\citep{tongyidr} remain constrained by their exclusive focus on \textit{textual web data}. This limitation overlooks the reality that professional and academic documents (\eg scientific papers, technical reports, financial reports, brochures, etc) are inherently multimodal, seamlessly integrating text with images, tables, equations, and charts~\citep{tang1996automatic,ye2023mplug}. Moreover, these systems lack the capability to process locally stored documents, forcing users to rely solely on publicly available web content rather than their own document repositories.

\begin{figure*}[t]
    \centering
    \includegraphics[width=0.98\linewidth]{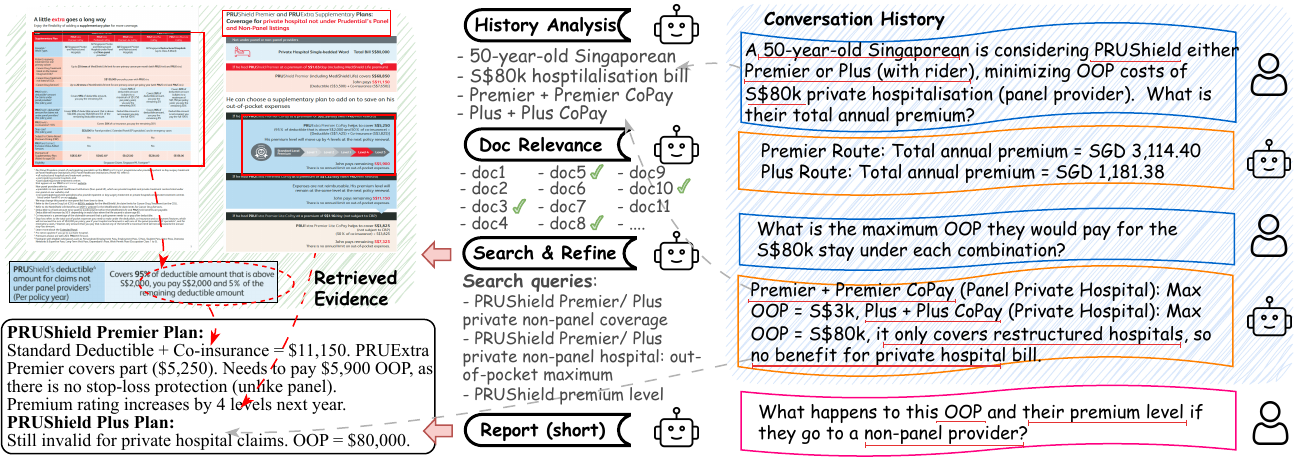}
    \vspace{-1.0em}
    \caption{An typical use case of multimodal doc deep research. The user asks a multi-hop question in the context of multi-turn conversations, where ground-truth evidence spans across multiple documents and modalities. The demo conversations, evidence, and answer are shortened for simplicity. Refer to more examples and evidence-chain annotations in Figure~\ref{fig:annotations}.}
    \vspace{-0.5em}
    \label{fig:task_view}
\end{figure*}

To facilitate multimodal document understanding, recent agentic RAG systems (\eg MDocAgent~\citep{han2025mdocagent} and M3DocRAG~\citep{cho2024m3docrag}) have emerged. However, these systems exhibit three critical limitations:
\textbf{(1) Inadequate multimodal parsing.} Current approaches rely on simplistic parsing strategies: either converting documents into OCR-text~\citep{wang2025vidorag} or treating them as raw page screenshots~\citep{faysse2024colpali, han2025mdocagent}. Both methods fail to preserve modality-specific characteristics and rich visual semantics inherent in charts, tables, figures, equations, and complex layouts.
\textbf{(2) Limited retrieval strategies.} Existing systems employ retrieval mechanisms based solely on OCR-extracted text chunks or full-page screenshots, hindering precise multimodal evidence localization. Different research tasks demand varying granularities: document-level summaries~\citep{dong2024-mc} for comparative analysis, fine-grained chunks~\citep{dong2023-chunks} for specific evidence extraction, or visual elements for multimodal understanding. Yet current systems lack mechanisms to dynamically select optimal retrieval strategies across modalities and granularity levels (\eg full-text, summary, page, chunk) based on query characteristics.
\textbf{(3) Absence of deep research capabilities.} Current systems are confined to single-round visual QA rather than supporting iterative, multi-step research workflows. Furthermore, no evaluation framework comprehensively assesses multimodal document understanding in realistic research scenarios: particularly those requiring large-scale document collections with annotated evidence chains, multi-hop reasoning, and multi-turn interactions (Figure~\ref{fig:task_view}). In contrast, existing document QA benchmarks focus primarily on single-document scenarios~\citep{dong2025mmdocir, ma2024mmlongbenchdoc} (Table~\ref{tab:dataset_compare}).

To address these challenges, we introduce \textbf{\mname}, a unified system that integrates sophisticated multimodal document parsing with multi-agent deep research capabilities. Our approach makes three key contributions:
\textbf{(1) Deep multimodal parsing ($\S$\ref{ssec:deep_parsing}).} We develop a comprehensive parsing framework that employs MinerU~\citep{wang2024mineru} for layout-aware document analysis, followed by intelligent chunking strategies and multimodal content processing. This framework creates multi-granular document representations that preserve both structural and semantic information, enabling efficient multimodal retrieval.
\textbf{(2) Systematic retrieval architecture ($\S$\ref{ssec:retrieval_arch}).} We conduct extensive experiments on multimodal retrieval paradigms, evaluating 5 text and 5 vision retrievers alongside various reranking and query extension strategies. Our analysis reveals the effectiveness-efficiency tradeoffs among text-only, vision-only, and hybrid approaches.
\textbf{(3) Deep research workflows ($\S$\ref{ssec:deep_research}).} \mname supports iterative, multi-step research workflows on local document collections through intelligent planning capabilities. The framework adaptively selects optimal retrieval granularities based on query characteristics and content types, enabling complex multi-hop reasoning across documents.

Beyond the system itself, we introduce \textbf{\dname ($\S$\ref{sec:dataset})}, a comprehensive benchmark for evaluating multimodal, multi-hop, multi-document, and multi-turn deep research capabilities. This benchmark provides expert-annotated question-answer pairs with complete evidence chains, enabling rigorous assessment of document understanding in realistic research scenarios. Experimental results demonstrate that \mname significantly outperforms existing approaches, establishing a new paradigm for multimodal document retrieval and deep research.
In summary, our contributions are threefold:
\textbf{(1) Deep multimodal parsing and retrieval framework}: We develop modular, plug-and-play components for multimodal document parsing and retrieval that can enhance any deep research system, featuring layout-aware analysis and multi-granular representation strategies.
\textbf{(2) \mname system}: We present the first unified deep research workflows, leveraging adaptive and granular retrieval, and iterative evidence refinement for multimodal documents understanding.
\textbf{(3) \dname benchmark}: We establish the first comprehensive benchmark for multimodal deep research, featuring large-scale document collections with annotated evidence chains for multi-hop, multi-document, and multi-turn evaluation.

\begin{figure*}[t]
    \centering
    \includegraphics[width=0.85\linewidth]{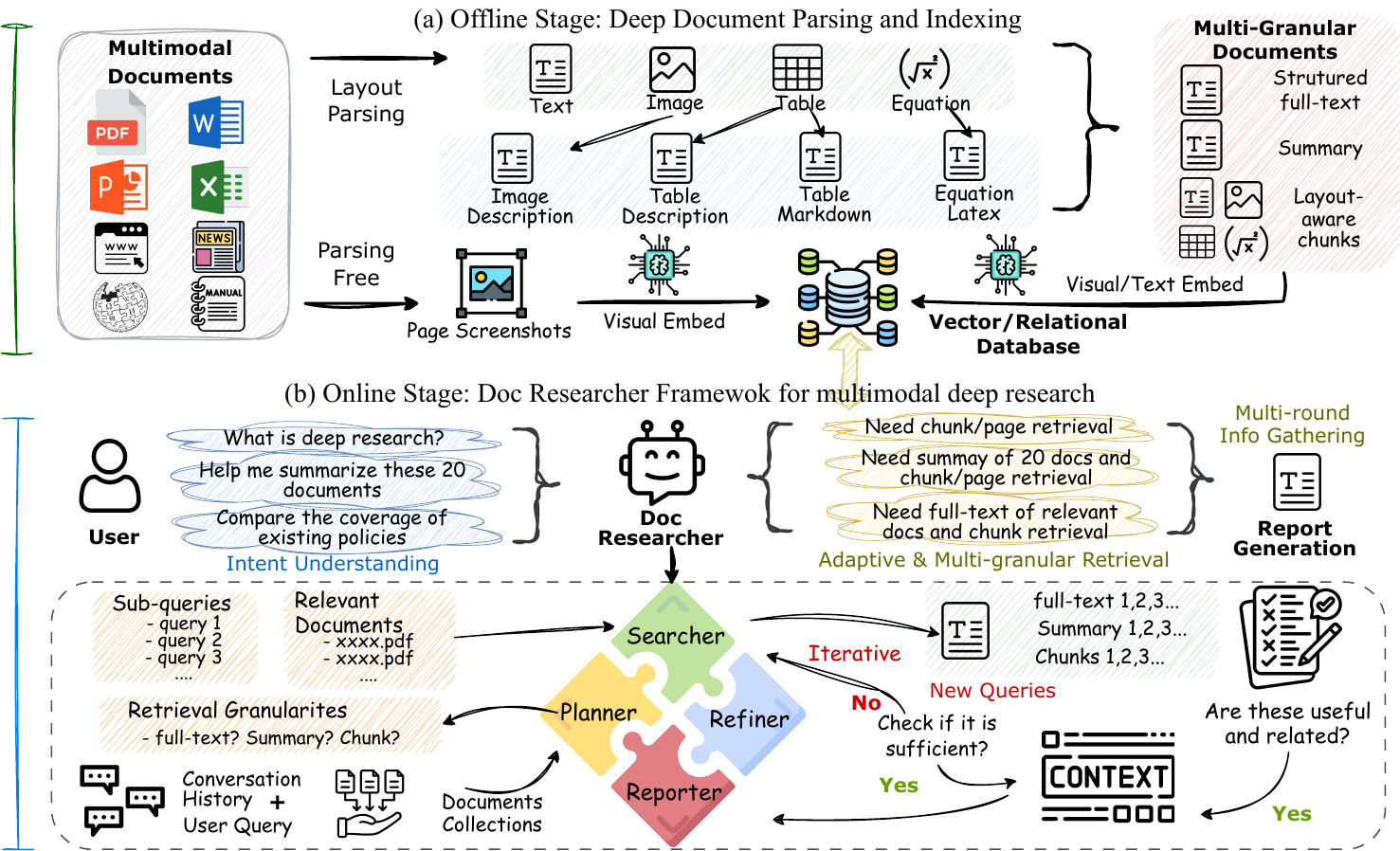}
    \vspace{-1.0em}
    \caption{\mname Architecture: (a) multimodal deep parsing and indexing, and (b) multimodal deep research.}
    \label{fig:deep_doc_researcher}
\end{figure*}

\section{Related Work}
\label{sec:related}
Our work builds upon three research directions that collectively define the landscape of multimodal document understanding.

\textit{\textbf{Document Parsing Strategies}.}
Document parsing approaches fall into three categories, each with distinct tradeoffs. \textbf{Shallow parsing} relies on OCR to extract text while discarding spatial and visual information~\citep{2007TessOverview}. \textbf{Deep parsing} systems like MinerU~\citep{wang2024mineru} preserve layout structure through bounding boxes, transcribe equations to LaTeX~\citep{wang2024unimernet}, and convert tables to structured formats~\citep{xia2024docgenome}. Recent VLM-based parsers~\citep{li2025monkeyocr, poznanski2025olmocr} further enhance extraction quality. \textbf{Parsing-free} approaches~\citep{ma2024dse, faysse2024colpali} bypass parsing entirely, processing documents as page screenshots. While avoiding parsing overhead, they sacrifice fine-grained localization and struggle with dense text. Our deep parsing framework synthesizes these approaches' strengths: preserving visual semantics like parsing-free methods while maintaining structural precision of deep parsing.

\textit{\textbf{Visual Document Understanding}.}
DocVQA benchmarks evolve from single-page tasks \citep{2020docvqa, mathew22infographicsvqa} to multi-hop reasoning across pages \citep{landeghem2023dude, tito2023mp-docvqa} and recently to long documents \citep{ma2024mmlongbenchdoc, zou2024docbench}. However, these benchmarks evaluate isolated capabilities: either VLM long-context processing (1-20 images) or single-document QA (50+ pages). Crucially, none assess multi-document synthesis, multi-turn interaction, or evidence chain annotation needed for research-level tasks. Hence, we propose \dname (Section~\ref{sec:dataset}) to address these gaps with comprehensive deep research evaluation.

\textit{\textbf{Document RAG Systems}.}
Due to limited context windows, DocRAG systems have merged that decompose understanding into retrieval and reasoning stages. Current systems~\citep{cho2024m3docrag, han2025mdocagent, dong2025mmdocrag} employ various strategies: M3DocRAG uses vision-only page retrieval, MDocAgent combines text chunks with screenshots, and VDocRAG~\citep{tanaka2025vdocrag} incorporates figure extraction. However, all perform single-round retrieval with fixed granularities, lacking the iterative refinement and adaptive strategies essential for complex research. \mname introduces the first deep research framework for multimodal documents, enabling multi-step investigation through dynamic granularity selection and progressive evidence synthesis.

\section{Method: \mname}
\label{sec:method}

\mname (in Figure~\ref{fig:deep_doc_researcher}) consists of two main modules: deep parsing and deep research.
The deep parsing pipeline transforms multimodal documents into structured, multi-granular, and searchable units while preserving multimodal semantics. 
The deep research adopts a multi-agent collaborative framework to solve complex questions with iterative search and refinement.

\subsection{Deep Multimodal Document Parsing}
\label{ssec:deep_parsing}

\textit{\textbf{Multi-modal Doc Parsing}.} 
Given a collection of multimodal documents $\mathcal{D}$, each document $d_i \in \mathcal{D}$ undergoes layout-aware parsing via MinerU~\citep{wang2024mineru} to extract structural elements $E_i = \{e_{i,j,k}\}$, where indices denote document $i$, page $j$, and reading sequence $k$. Each element is pair with its layout type (text, table, figure, equation), and bounding box coordinates (Figure~\ref{fig:layout_chunk}a). 
For computational efficiency, we perform one-time textual conversion of visual elements. Tables and figures are processed through Qwen2.5-VL~\citep{bai2025qwen25vl} to generate both coarse-grained summaries for contextual understanding and fine-grained descriptions for detailed content capture. Equations are converted to LaTeX format using UniMERNet~\citep{wang2024unimernet}. This preprocessing strategy enables efficient retrieval and reasoning operations while avoiding repeated multimodal token processing.

\textit{\textbf{Multi-granular Doc Chunking}.} 
While individual layout elements $E_i$ preserve structural integrity, they often lack sufficient context. We address this through layout-aware chunking that merges text elements within section boundaries, subject to maximum length constraints (Figure~\ref{fig:layout_chunk}b). This strategy maintains semantic coherence while preserving full traceability via page IDs and bounding box coordinates for precise localization and citation.
We construct 4 different granularity levels $G = \{\text{chunk}, \text{page}, \text{full}, \text{summary} \}$: (1) chunks created through layout-aware merging, (2) pages by either combining all elements $e_{i,j,\cdot}$ within page $j$, or raw page screenshots, (3) full-text containing all elements within document $d_i$, and (4) document summary is obtained by summarizing full-text of document $d_i$ using LLMs.

\begin{table*}[t]
\centering
\renewcommand{\arraystretch}{0.85}
\resizebox{0.99\linewidth}{!}{%
\begin{tabular}{l@{\hskip 1pt}|c@{\hskip 3.5pt}c@{\hskip 3.5pt}c@{\hskip 5.5pt}c|c|c@{\hskip 2pt}|c@{\hskip 2pt}|c@{\hskip 2pt}|c@{\hskip 6pt}c@{\hskip 6pt}c@{\hskip 5pt}c@{\hskip 3pt}c@{\hskip 3pt}c@{\hskip 4pt}c@{\hskip 2pt}}
    \toprule
    \multirow{2}{*}{Benchmarks}  & \multicolumn{4}{c|}{Question of multi-} & \multirow{2}{*}{QA} & \multirow{2}{*}{Annotation Type} & \multirow{2}{*}{Domain} & \multirow{2}{*}{lang'} & \multicolumn{7}{c}{Retrieval Annotation} \\
    & hop & modal & doc & turn &  &  &  &  & doc & \#doc & page & \#page & layout & \#lay & subquery \\
    \midrule

    ViDoRe~\citep{faysse2024colpali} & \xmark & \cmark & \xmark & \xmark & 3,810 & Existing+Claude3 & Multiple & 2 & \xmark & 1.0 & \cmark & 1.0 & \xmark & - & \xmark \\
    
    MMLongBench-Doc~\citep{ma2024mmlongbenchdoc} & \cmark & \cmark & \xmark & \xmark & 1,082 & Expert & 7 domains & 1 & \xmark & 1.0 & \cmark & 1.2 & \xmark & - & \xmark \\

    DocBench~\citep{zou2024docbench} & \cmark & \cmark & \xmark & \xmark & 1,102 & GPT4 + Human & 5 domains & 1 & \xmark & 1.0 & \xmark & - & \xmark & - & \xmark \\

    M3SciQA~\citep{li2024m3sciqa} & \xmark & \cmark & \xmark & \xmark & 1,452 & GPT-4 + Human & NLP papers & 1 & \cmark & 2.0 & \xmark & - & \xmark & - & \xmark \\

    M3DocVQA~\citep{cho2024m3docrag} & \xmark & \cmark & \xmark & \xmark & 2,441 & Human & Wikipedia & 1 &  \xmark & 1.0 & \cmark & 1.4 & \xmark & - & \xmark \\

    ViDoSeek~\citep{wang2025vidorag} & \cmark & \cmark & \xmark & \xmark & 3,162 & Human & Slide & 1 & \xmark & 1.0 & \cmark & 1.0 & \xmark & - & \xmark \\

    MMDocIR~\citep{dong2025mmdocir} & \cmark & \cmark & \xmark & \xmark & 1,658 & Expert &  10 domains & 1 & \xmark & 1.0 & \cmark & 1.5 & \cmark & 1.8 & \xmark \\

    ViDoRe-v2~\citep{macé2025vidorev2}  & \cmark & \cmark & \cmark & \xmark  & 271 & GPT-4o + Human & ESG/Bio/Eco & 4 & \xmark & 1.0 & \cmark & 5.1 & \xmark & - & \xmark \\
    
    MMDocRAG~\citep{dong2025mmdocrag} & \cmark & \cmark & \xmark & \xmark  & 4,055 & GPT-4o + Human & 10 domains & 1 & \xmark & 1.0 & \cmark & 1.4 & \cmark & 1.7 & \xmark \\
    
    Double-Bench~\citep{shen2025doublebench} & \cmark & \cmark & \xmark & \xmark & 5,168 & KG+GPT-4o+Human & Multiple & 6 & \xmark & 1.0 & \cmark & 5.9 & \xmark & - & \cmark \\

    \midrule    
    \textbf{\dname} & \cmark & \cmark & \cmark & \cmark & 158 & Expert & 4 domains & 2 & \cmark & 3.8 & \cmark & 7.0 & \cmark & 14.8 & \cmark \\

    \bottomrule

\end{tabular}}
\caption{\dname compared to other Document RAG/VQA dataset.}
\vspace{-2em}
\label{tab:dataset_compare}
\end{table*}

\subsection{Multimodal Retrieval Architecture}
\label{ssec:retrieval_arch}

\textit{\textbf{Vision-only Retrieval}.}
Vision-only retrieval operates at page granularity using raw screenshots as encoded passages. This approach eliminates parsing overhead, offering high efficiency without OCR or layout analysis. However, it faces challenges in representing complex interleaved text-vision content through dense vectors.
For high-resolution pages, multi-vector vision retrievers can produce thousands of vectors, as each small image patch (typically 16x16 or 14x14 pixels) correspond to a vector. Additionally, vision retrievers rely on parameter-heavy VLMs (often >3B parameters) with deeper architectures, increasing computational costs.

\textit{\textbf{Text-only Retrieval}.}
Text-only retrieval supports both page and chunk granularities, encoding OCR-extracted text, text chunks, and VLM-generated descriptions of visual elements. This approach leverages lightweight models (often <1B parameters) that excel at encoding text-intensive documents. However, generating VLM descriptions incurs preprocessing costs and may lose critical visual semantics during textual conversion.

\textit{\textbf{Hybrid Retrieval}.}
Hybrid retrieval combines strengths of both paradigms by directly encoding visual chunks and page screenshots without intermediate textual conversion, preserving rich visual information while maintaining text understanding capabilities. This approach eliminates information loss from OCR or VLM description but requires more computational resources during inference. The trade-off between retrieval quality and efficiency depends on specific application requirements and document characteristics.

\subsection{Multimodal Deep Research}
\label{ssec:deep_research}

\textit{\textbf{Strategic Document Filtering and Adaptive Processing}.} 
Given a query $q_i$ with dialog history $h_i$, the Planner agent addresses computational efficiency in large-scale document analysis through intelligent document selection and determining retrieval granularity. The agent analyzes $(q_i, h_i)$ to produce three outputs: (i) a filtered document subset $\mathcal{D}' \subseteq \mathcal{D}$ by matching query semantics against document summaries, reducing search space by 60-80\% while maintaining high recall; (ii) optimal retrieval granularity $\theta \in G =\{\text{summary}, \text{chunk}, \text{full}, \text{summary}\}$ based on query characteristics; and (iii) refined sub-queries at timestamp $Q_t=\{\tilde{q}_1, \ldots, \tilde{q}_n\}$ that decompose complex questions or explore in new search directions. This adaptive strategy leverages the multi-granular representations from our parsing framework ($\S$\ref{ssec:deep_parsing}). For instance, broad contextual queries activate summary mode to access document overviews directly, while technical queries that require specific evidence trigger targeted chunk extraction through iterative refinement.

\textit{\textbf{Iterative Search-Refine Loop}.} 
The core research process operates through dynamic collaboration between Searcher and Refiner. For each sub-query $\tilde{q}_k \in Q_t$, the system executes an iterative loop:
\begin{align}
\mathcal{R}_t &= \text{Search}(\tilde{q}_k, \mathcal{D}', \theta), \quad \mathcal{R}_t^* = \text{Refine}(\mathcal{R}_t, \tilde{q}_k) \\
\sigma_t &= \text{Evaluate}(\mathcal{R}_1^* \cup \ldots \cup \mathcal{R}_t^*, q_i)
\end{align}
where $\mathcal{R}_t$ represents content retrieved at iteration $t$ using our multimodal retrieval architecture ($\S$\ref{ssec:retrieval_arch}), $\mathcal{R}_t^*$ is the refined relevant subset after deduplication and relevance filtering, and $\sigma_t$ measures information sufficiency. 
The loop continues until either sufficiency threshold is met ($\sigma_t \geq \tau$) or maximum iterations reached ($t = T_{\max}$), balancing thoroughness with computational efficiency. If a new loop is required, the collection of subqueries is updated concurrently ($Q_t\rightarrow Q_{t+1}$) for the next iteration.

\textit{\textbf{Multimodal Report Generation}.} 
The Reporter synthesizes accumulated evidence $\mathcal{R}^* = \bigcup_t \mathcal{R}_t^*$ into a comprehensive response $\hat{a} = \text{Report}(q_i, \mathcal{R}^*, \mathcal{C})$, where $\mathcal{C}$ contains citation metadata (page IDs and bounding boxes coordinates) from our parsing framework. By analyzing query intent and retrieved information, the Reporter generates interleaved text-image outputs that directly incorporate relevant visual elements (\eg tables, figures, and charts) alongside textual explanations. This supports the notion that ``\emph{a single image is worth a thousand words}''. Moreover, the multimodal composition enhances both answer quality and verifiability, as users can directly verify claims to precise document locations via embedded multimodal citations. The approach proves particularly effective for complex analytical queries requiring evidence synthesis across diverse multimodal sources.

\section{\dname Benchmark}
\label{sec:dataset}

\subsection{Motivation of \dname}

Existing evaluation benchmarks fail to capture the complexity of multimodal document deep research. A comprehensive evaluation must assess four critical dimensions:
\textbf{(1) Multi-hop reasoning}: Complex questions requiring sophisticated reasoning chains across evidence.
\textbf{(2) Multi-modal integration}: Questions demanding information from at least two modalities (text, tables, images).
\textbf{(3) Multi-document synthesis}: Questions relying evidence from multiple documents, where each document provides unique, essential information. Distractor documents with topically related content increase retrieval difficulty.
\textbf{(4) Multi-turn interaction}: Dialogical contexts reflecting real-world research sessions, where current queries require information and disambiguation of prior exchanges.

Table~\ref{tab:dataset_compare} reveals critical gaps in existing benchmarks. While most DocVQA/RAG datasets incorporate multi-hop and multi-modal evidence, only one includes genuine multi-document reasoning.\footnote{Datasets that merely add noise documents to single-document VQA questions without proper de-contextualization do not constitute true multi-document settings.} Most critically, the limited evidence scope (averaging fewer than 2 pages per question) indicates relatively simple and localized queries rather than comprehensive research tasks. Also, only 2 existing benchmarks offer layout-level evidence localization. Furthermore, no existing benchmark provides multi-turn conversational evaluation.
To address these limitations, we introduce \textbf{\dname}, a benchmark specifically designed for \textbf{M}ulti-modal, \textbf{M}ulti-hop, \textbf{M}ulti-document, and \textbf{M}ulti-turn deep research evaluation. \dname features expert-annotated question-answer pairs with complete evidence chains, enabling rigorous assessment of document understanding capabilities in realistic research scenarios.

\subsection{Domain and Document Collection}

We select four distinct domains (research, insurance, education, and finance) based on three criteria: (1) \textbf{Multimodal richness}: Each domain features essential visual elements (tables, charts, figures) integrated with text; (2) \textbf{Reasoning complexity}: Questions require diverse capabilities from technical comprehension to numerical analysis; (3) \textbf{Practical relevance}: These represent real-world scenarios where professionals routinely conduct document research.
We collect documents from authoritative sources to ensure quality and representativeness:
\begin{itemize}[leftmargin=*, itemsep=0em, topsep=-0.1em]
    \item \textbf{Research}: We collect 166 computer science papers from arXiv and top-tier conferences (ICLR, NeurIPS, ICML, ACL, CVPR), averaging 26.7 pages and 21.8k words in length.
    
    \item \textbf{Insurance}: We gather 43 brochures and policy documents from major insurance insurers , averaging 11.0 pages and 5.5k words.
    
    \item \textbf{Education}: We collect 56 program materials (averaging 7.5 pages and 3.9k words) from prestigious universities, including merged program webpages and separate supplementary documents.
    
    \item \textbf{Finance}: We obtain 38 market research reports (22.2 pages and 11.2k words on average), covering monthly and half-yearly trends across A-shares, Hong Kong, and NASDAQ.
\end{itemize}

\subsection{Annotation Process}
\label{ssec:annotation_process}

Our annotation process employs PhD/Master-level researchers with extensive research experience to ensure benchmark quality. Beyond basic question-answer pairs, we annotate six critical dimensions that enable comprehensive evaluation of deep research systems:
\begin{itemize}[leftmargin=*, itemsep=0em, topsep=-0.1em]
    \item \textbf{Hard Negative Documents}: We carefully curate documents that share topical similarity with target questions while lacking essential information. This tests systems' ability to distinguish truly relevant sources from plausible distractors—a critical capability for effective document filtering in large collections.
    
    \item \textbf{Retrieval Granularity}: We label optimal information extraction levels (full-text, summary, page, chunk) based on query characteristics. For example, single-document summarization requires full-text retrieval, while multi-document synthesis necessitates summary or chunk retrieval due to context constraints. This enables evaluation of adaptive granularity selection.
    
    \item \textbf{Fine-grained Layout}: Following MMDocIR~\citep{dong2025mmdocir}, we annotate both relevant pages and bounding boxes within those pages, enabling precise evaluation at element levels.
    
    \item \textbf{Subquery Decomposition}: We provide ground-truth intermediate queries that reveal optimal decomposition strategies for complex questions. Since deep research systems typically break down questions iteratively, these annotations enable evaluation of query planning and refinement capabilities.
    
    \item \textbf{Answer Verification Checklists}: For lengthy responses where objective evaluation proves challenging, we develop structured checklists with clear criteria for assessing answer completeness and accuracy, enabling systematic LLM-based evaluation.
    
    \item \textbf{Bilingual Support}: All annotators are proficient in English and Chinese, enabling annotation of questions and documents in both languages to evaluate multilingual system capabilities.
\end{itemize}

To facilitate understanding on \dname, we demonstrate one annotation example in Figure~\ref{fig:task_view} and two examples in Figure~\ref{fig:annotations}.

\subsection{Benchmark Statistics}
\label{ssec:benchmark_stats}

Table~\ref{tab:dataset_main} presents statistics of \dname's 158 expert-annotated questions, demonstrating its unique evaluation capabilities:
\textbf{(1) Document-scale Complexity.} 
Questions require processing 12.7 documents (306.4 pages) on average, up to 42 documents (836 pages) in challenging cases. Systems must identify 3.8 relevant documents from extensive collections, testing critical document filtering capabilities underexplored in existing benchmarks.
\textbf{(2) Multimodal and Complex Evidence Requirements.}
112 questions require multiple modalities, with balanced evidence distribution across text (40.9\%), tables (30.2\%), and images (28.9\%). Fine-grained evidence annotations average 7.0 pages and 14.8 layout elements per question, indicating the complexity of \dname.
\textbf{(3) Multi-turn Research Depth.}
58 questions occur in multi-turn conversations, capturing iterative research workflows. Questions average 3.9 search clauses and answers range from 59.7 to 272.0 words, testing adaptive query planning and output granularity.

\section{Experiment}
\label{sec:experiment}

\begin{table*}[ht]
\renewcommand{\arraystretch}{0.6}
\centering
\resizebox{0.9\linewidth}{!}{
\begin{tabular}{lll|c@{\hskip 2.5pt}c@{\hskip 2.5pt}c|ccc|cc}
    \toprule
    \multicolumn{3}{c|}{\multirow{2}{*}{\textbf{Model}}} & \textbf{Para-} & \multicolumn{2}{c|}{\textbf{Embed}} & \multicolumn{3}{c|}{\textbf{M4DocBench}} & \multicolumn{2}{c}{\textbf{MMDocIR}} \\
     & & & \textbf{meter} & type & dimension & doc & page & layout & page & layout \\
    \midrule
    
    \parbox[t]{1.5mm}{\multirow{17}{*}{\rotatebox[origin=c]{90}{Recall@$k=10$}}} 
    & \parbox[t]{1.0mm}{\multirow{5}{*}{\rotatebox[origin=c]{90}{Text}}} 
    & BM25 & - & sparse & -                     & 51.5 / 59.3 & 22.1 / 25.6 & 15.8 / 17.1 & 60.0 / 65.0 & 45.1 / 48.8 \\
    && E5 & 0.34B & dense & 1,024               & 65.7 / 75.4 & 29.9 / 31.8 & 18.0 / 18.0 & 54.4 / 61.3 & 38.0 / 44.8 \\
    && BGE-M3 & 0.56B & dense & 1,024           & 75.4 / \textbf{81.4} & 33.3 / 40.0 & 19.2 / 20.8 & 58.0 / 61.9 & 41.3 / 44.9 \\
    && Qwen3-embedding & 0.60B & dense & 1,024   & 73.6 / 78.8 & 34.6 / 40.6 & 20.1 / \textbf{23.5} & 58.6 / 62.6 & 42.6 / 45.6 \\
    && Multi+reranking$^\ast$ & 1.5B & hyrbid & - &\textbf{\textcolor{purple}{78.7}} / 80.3 & \textbf{\textcolor{purple}{39.2}} / \textbf{\textcolor{purple}{40.9}} & \textbf{23.1} / 21.1 & \textbf{65.5} / \textbf{66.7} & \textbf{48.6} / \textbf{49.4} \\
    
    \cmidrule{3-11}

    & \parbox[t]{2.0mm}{\multirow{5}{*}{\rotatebox[origin=c]{90}{Vision}}} 
    & DSE$_{\mathrm{wiki-ss}}$ & 4.15B & dense & 3,072       & 34.8 / 42.4 & 11.1 / 11.7 & - & 68.4 / 68.7 & - \\
    && ColPali & 2.92B & multi & $N_{\mathrm{tok}}\times$128 & 47.7 / 50.0 & 18.1 / 17.9 & - & 75.4 / 73.9 & - \\
    && ColQwen & 2.21B & multi & $N_{\mathrm{tok}}\times$128 & 49.8 / 55.8 & 25.5 / 24.2 & - & 77.4 / 75.4 & - \\
    && Jina-embedding-v4 & 3.75B & dense & 2,048             & 52.9 / 56.3 & 23.6 / 22.9 & - & 74.4 / 71.5 & - \\
    && Jina-embedding-v4 & 3.75B & multi & $N_{\mathrm{tok}}\times$128 & \textbf{54.3} / \textbf{58.7} & \textbf{25.8} / \textbf{26.7} & - & \textbf{80.8} / \textbf{79.4} & - \\
    \cmidrule{3-11}

    & \parbox[t]{2.0mm}{\multirow{3}{*}{\rotatebox[origin=c]{90}{Hyb'd}}} 
    & Jina-embedding-v4 & 3.75B & dense & 2,048         & 74.0 / 79.4 & \textbf{39.1} / \textbf{40.5} & \textbf{\textcolor{purple}{28.7}} / \textbf{\textcolor{purple}{27.9}} & 83.4 / \textbf{\textcolor{purple}{84.0}} & 66.1 / 65.8 \\
    && Qwen3 + Jina & 0.60/3.75B & dense & 1,024/2,048  & 74.0 / 79.6 & 37.5 / 37.5 & 28.1 / 26.9 & 82.9 / 83.3 & 65.4 / 64.7 \\
    && BGE-M3 + Jina & 0.56/3.75B & dense & 1,024/2,048 & \textbf{77.1} / \textbf{\textcolor{purple}{81.8}} & 38.4 / 39.1 & 28.1 / 27.5 & \textbf{\textcolor{purple}{83.7}} / 83.9 & \textbf{\textcolor{purple}{66.2}} / \textbf{\textcolor{purple}{66.0}} \\

    \midrule
    
    \parbox[t]{1.5mm}{\multirow{17}{*}{\rotatebox[origin=c]{90}{Recall@$k=15$}}} 
    & \parbox[t]{1.0mm}{\multirow{5}{*}{\rotatebox[origin=c]{90}{Text}}} 
    & BM25 & - & sparse & -                     & 58.1 / 66.3 & 28.0 / 32.6 & 20.3 / 22.9 & 62.4 / 68.0 & 46.8 / 52.2 \\
    && E5 & 0.34B & dense & 1,024               & 75.4 / 81.9 & 35.3 / 41.7 & 22.1 / 25.2 & 58.0 / 64.4 & 41.2 / 48.0 \\
    && BGE-M3 & 0.56B & dense & 1,024           & 83.3 / \textbf{\textcolor{purple}{87.6}} & 41.2 / 46.8 & 23.8 / 24.9 & 61.4 / 64.8 & 44.4 / 47.9 \\
    && Qwen3-embedding & 0.60B & dense & 1,024   & 81.1 / 86.8 & 42.1 / \textbf{\textcolor{purple}{49.9}} & 25.4 / 27.4 & 61.4 / 65.3 & 45.2 / 48.3 \\
    && Multi+reranking$^\ast$ & 1.5B & hyrbid & -   & \textbf{83.6} / 87.2 & \textbf{44.8} / 49.4 & \textbf{27.4} / \textbf{27.9} & \textbf{67.9} / \textbf{70.1} & \textbf{51.0} / \textbf{53.0} \\
    \cmidrule{3-11}  

    & \parbox[t]{2.0mm}{\multirow{5}{*}{\rotatebox[origin=c]{90}{Vision}}} 
    & DSE$_{\mathrm{wiki-ss}}$ & 4.15B & dense & 3,072       & 45.2 / 53.9 & 18.1 / 18.5 & - & 78.0 / 78.5 & - \\
    && ColPali & 2.92B & multi & $N_{\mathrm{tok}}\times$128 & 55.7 / 61.7 & 24.7 / 24.9 & - & 82.0 / 80.9 & - \\
    && ColQwen & 2.21B & multi & $N_{\mathrm{tok}}\times$128 & 60.6 / 69.4 & 33.8 / 33.3 & - & 83.1 / 82.9 & - \\
    && Jina-embedding-v4 & 3.75B & dense & 2,048             & 64.7 / 70.9 & 32.2 / 33.0 & - & 80.3 / 79.4 & - \\
    && Jina-embedding-v4 & 3.75B & multi & $N_{\mathrm{tok}}\times$128 & \textbf{66.0} / \textbf{74.4} & \textbf{35.2 }/ \textbf{38.8} & - & \textbf{85.8} / \textbf{ 84.7} & - \\
    \cmidrule{3-11}

    & \parbox[t]{2.0mm}{\multirow{3}{*}{\rotatebox[origin=c]{90}{Hyb'd}}} 
    & Jina-embedding-v4 & 3.75B & dense & 2,048         & 83.6 / \textbf{\textcolor{purple}{87.6}} & \textbf{\textcolor{purple}{49.2}} / \textbf{\textcolor{purple}{49.9}} & \textbf{\textcolor{purple}{35.6}} / \textbf{\textcolor{purple}{35.2}} & 87.9 / 87.9 & 70.6 / \textbf{\textcolor{purple}{70.5}} \\
    && Qwen3 + Jina & 0.60/3.75B & dense & 1,024/2,048  & 83.9 / 86.5 & 46.0 / 47.3 & 33.5 / 33.7 & 87.0 / 87.3 & 69.7 / 69.3 \\
    && BGE-M3 + Jina & 0.56/3.75B & dense & 1,024/2,048 & \textbf{\textcolor{purple}{84.1}} / 87.0 & 44.9 / 47.1 & 33.4 / 34.1 &\textbf{ \textcolor{purple}{88.5}} / \textbf{\textcolor{purple}{88.1}} & \textbf{\textcolor{purple}{70.9}} / 70.4  \\

    \midrule

    \parbox[t]{1.5mm}{\multirow{17}{*}{\rotatebox[origin=c]{90}{Recall@$k=20$}}} 
    & \parbox[t]{1.0mm}{\multirow{5}{*}{\rotatebox[origin=c]{90}{Text}}} 
    & BM25 & - & sparse & -                     & 61.3 / 69.3 & 30.2 / 36.3 & 22.1 / 24.8 & 64.5 / 65.3 & 48.9 / 49.5 \\
    && E5 & 0.34B & dense & 1,024               & 79.4 / 85.1 & 39.4 / 47.5 & 25.6 / 29.0 & 60.1 / 66.4 & 43.3 / 49.7 \\
    && BGE-M3 & 0.56B & dense & 1,024           & 86.6 / 89.9 & 46.8 / 52.3 & 27.3 / 28.7 & 62.8 / 67.0 & 45.5 / 49.8 \\
    && Qwen3-embedding & 0.60B & dense & 1,024   & 84.9 / \textbf{\textcolor{purple}{90.6}} & 47.2 / 56.0 & 30.5 / 30.4 & 63.5 / 66.9 & 47.0 / 50.2 \\
    && Multi+reranking$^\ast$ & 1.5B & hyrbid & -   & \textbf{87.8} / 90.3 & \textbf{50.6} / \textbf{56.3} & \textbf{31.9} / \textbf{32.3} & \textbf{70.1} / \textbf{71.7} & \textbf{53.5} / \textbf{54.9} \\
    \cmidrule{3-11}  

    & \parbox[t]{2.0mm}{\multirow{5}{*}{\rotatebox[origin=c]{90}{Vision}}} 
    & DSE$_{\mathrm{wiki-ss}}$ & 4.15B & dense & 3,072       & 52.5 / 59.5 & 23.3 / 23.5 & - & 83.1 / 83.8 & - \\
    && ColPali & 2.92B & multi & $N_{\mathrm{tok}}\times$128 & 61.5 / 67.8 & 29.5 / 30.0 & - & 86.0 / 85.6 & - \\
    && ColQwen & 2.21B & multi & $N_{\mathrm{tok}}\times$128 & 67.8 / 77.0 & 40.0 / 40.9 & - & 86.8 / 86.9 & - \\
    && Jina-embedding-v4 & 3.75B & dense & 2,048             &\textbf{73.0} / 78.2 & 39.7 / 41.3 & - & 84.6 / 84.0 & - \\
    && Jina-embedding-v4 & 3.75B & multi & $N_{\mathrm{tok}}\times$128 & 70.9 / \textbf{80.5} & \textbf{42.1} / \textbf{46.3} & - & \textbf{88.9} / \textbf{88.4} & - \\
    \cmidrule{3-11}

    & \parbox[t]{2.0mm}{\multirow{3}{*}{\rotatebox[origin=c]{90}{Hyb'd}}} 
    & Jina-embedding-v4 & 3.75B & dense & 2,048         &\textbf{\textcolor{purple}{88.6}} / 89.9 & \textbf{\textcolor{purple}{55.5}} / \textbf{\textcolor{purple}{56.5}} & \textbf{\textcolor{purple}{40.9}} / \textbf{\textcolor{purple}{40.6}} & 90.0 / \textbf{\textcolor{purple}{90.7}} & 73.4 / 73.6 \\
    && Qwen3 + Jina & 0.60/3.75B & dense & 1,024/2,048  & 87.2 / \textbf{90.1} & 53.1 / 54.5 & 39.4 / 39.7 & 89.8 / 89.7 & 72.2 / 72.5 \\
    && BGE-M3 + Jina & 0.56/3.75B & dense & 1,024/2,048 & 88.2 / 89.9 & 50.9 / 53.8 & 38.4 / 39.5 & \textbf{\textcolor{purple}{90.7}} / \textbf{\textcolor{purple}{90.7}} & \textbf{\textcolor{purple}{73.7}} / \textbf{\textcolor{purple}{73.9}} \\

    \bottomrule
\end{tabular}}
\caption{Main results for evaluating retrieval performance. Each score pair ($s_1$/$s_2$) are calculated using original question / decomposed sub-queries for retrieval. For $^\ast$ retrieval, we use multiple text retrievers (BM25, BGE-M3, Qwen3-embedding) to get chunks and subsequently reranked by Qwen3-reranker. The best results of text/vision/hybrid are in \textbf{boldface}, and the overall best results across text\&vision\&hybrid are further \textcolor{purple}{colored}.}
\vspace{-2em}
\label{tab:retrieval_main}
\end{table*}


\subsection{Evaluation Metrics and Benchmarks}

\textit{\textbf{\quad Multimodal Retrieval Evaluation}.} 
We evaluate retrieval performance at three granularities aligned with our retrieval framework ($\S$\ref{ssec:retrieval_arch}): document, page, and layout. For each granularity, we compute recall@k by comparing retrieved passages\footnote{Passages denote the minimal retrieval units: chunks for parsed methods, pages for parsing-free approaches.} against gold annotations. Document and page recall are computed using exact ID matching. For chunk-level evaluation, following \citet{dong2025mmdocir}, we calculate recall based on bounding box overlap between retrieved and gold-standard layouts. Note that parsing-free methods cannot be evaluated at chunk granularity.

\textit{\textbf{Deep Research Evaluation}.} 
We evaluate the complete deep research workflow ($\S$\ref{ssec:deep_research}) across three dimensions:
\textbf{(1) Document Selection}: We assess the Planner's ability to filter relevant documents from noisy collections, computing precision, recall, and F1 scores against gold document sets.
\textbf{(2) Agentic Retrieval}: After iterative search-refine loops, we evaluate the quality of accumulated relevant passages using the same metrics as multimodal retrieval, but with variable passage counts reflecting the system's adaptive selection.
\textbf{(3) Answer Accuracy}: Using LLM-as-judge with annotated checklists $\mathbf{K}=\{K_1,K_2,...,K_m\}$, we verify whether each atomic fact appears in response $\hat{a}$. Answers are marked correct only when all checklist items are satisfied.

\textit{\textbf{Evaluation Benchmarks}.} 
We employ two benchmarks: (1) \textbf{\dname} ($\S$\ref{sec:dataset}) for comprehensive evaluation of both multimodal retrieval and deep research capabilities across multi-hop, multi-modal, multi-document, and multi-turn scenarios; (2) \textbf{MMDocIR}~\citep{dong2025mmdocir} containing 1,658 single-document VQA questions for focused evaluation of page and chunk-level retrieval performance.\footnote{\url{https://github.com/MMDocRAG/MMDocIR}}

\begin{table*}[ht]
\centering
\renewcommand{\arraystretch}{0.75}
\setlength{\tabcolsep}{4.5pt}
\resizebox{1.0\linewidth}{!}{%
\begin{tabular}{l|l|c|c|c|cc|cccc|ccc|cccc}
    \toprule
    \multirow{2}{*}{Method} & \multirow{2}{*}{Backbone} & \multirow{2}{*}{Retriever} & Parsing & \multicolumn{7}{c|}{Accuracy} & \multicolumn{3}{c|}{Doc Selection} & \multicolumn{4}{c}{Retrieval Recall} \\
    & & & Level & All & en & zh & ins' & res' & edu' & fin' & Rec & Prec & F$_1$ & Doc & Page & Lay' & \#psg \\
    \midrule

    Direct & Qwen3-32B & - & - & 7.0 & 3.8 & 10.1 & 8.3 & 10.3 & 0.0 & 5.6 & - & - & - & - & - & - & -  \\
    Direct & Qwen3-235B & - & - &  5.1 & 5.1 & 5.1 & 4.2 & 7.4 & 3.3 & 2.8 & - & - & - & - & - & - & -  \\
    Direct & DeepSeek-R1 & - & - & 10.1 & 13.9 & 6.3 & 20.8 & 11.8 & 10.0 & 0.0  & - & - & - & - & - & - & -  \\
    Long-context & Qwen3-32B & - & Shallow & 13.3 & 19.0 & 7.6 & 25.0 & 20.6 & 0.0 & 2.8 & - & - & - & - & - & - & -  \\
    Long-context & Qwen3-235B & - & Shallow & 8.9 & 5.1 & 12.7 & 12.5 & 14.7 & 3.3 & 0.0 & - & - & - & - & - & - & -  \\
    Long-context & DeepSeek-R1 & - & Shallow & 29.1 & 29.1 & 29.1 & 20.8 & 47.1 & 20.0 & 8.3 & - & - & - & - & - & - & -  \\
    Long-context & Qwen3-32B & - & Deep & 13.3 & 12.7 & 13.9 & 20.8 & 22.1 & 3.3 & 0.0 & - & - & - & - & - & - & -  \\
    Long-context & Qwen3-235B & - & Deep & 11.4 & 7.6 & 15.2 & 16.7 & 19.1 & 3.3 & 0.0 & - & - & - & - & - & - & -  \\
    Long-context & DeepSeek-R1 & - & Deep & 31.7 & 34.2 & 29.1 & 33.3 & 45.6 & 30.0 & 5.6 & - & - & - & - & - & - & -  \\

    \midrule
    
    MDocAgent & InternVL3.5-38B & Hyribd (Multi)$^\heartsuit$ & Shallow & 15.8 & 19.0 & 12.7 & 20.8 & 23.5 & 6.7 & 5.6 & - & - & - & 74.22 & 37.54 & - & 10\\  
    M3DocRAG & InternVL3.5-38B & Vision (ColPali) & Free & 7.0 & 10.1 & 3.8 & 4.2 & 2.9 & 26.7 & 0.0 & - & - & - & 69.58 & 34.35 & - & 10   \\
    Colqwen-gen & InternVL3.5-38B  & Vision (ColQwen) & Free & 5.7 & 6.3 & 5.1 & 0.0 & 8.8 & 10.0 & 0.0 & - & - & - & 74.38 & 46.29 & - & 10  \\
    \midrule

    \mname & Qwen3-32B$^\clubsuit$ & Vision (Jina) & Free  & 36.7 & 34.2 & 39.2 & 20.8 & 52.9 & 40.0 & \textbf{\textcolor{purple}{13.9}} & - & - & - & 88.9 & \underline{61.3} & - & 17.9 \\
    \mname & Qwen3-235B$^\clubsuit$ & Vision (Jina) & Free  & 36.7 & 36.7 & 36.7 & 29.2 & 50.0 & 40.0 & \textbf{\textcolor{purple}{13.9}} & - & - & - & \underline{89.0} & 61.0 & - & 17.9 \\
    \mname & DeepSeek-R1$^\clubsuit$  & Vision (Jina) & Free & 43.7 & 45.6 & 41.7 & 37.5 & 61.8 & 43.3 & \textbf{\textcolor{purple}{13.9}} & - & - & - & \textbf{\textcolor{purple}{90.6}} & \textbf{\textcolor{purple}{63.6}} & - & 17.4 \\

    \mname & Qwen3-32B & Text (Multi)$^\spadesuit$ & Shallow & 34.2  & 37.9 & 30.4 & 12.5 & 55.9 & 36.7 & 5.6 & - & - & - & 88.2 & 59.8 & - & 14.7 \\
    \mname & Qwen3-235B & Text (Multi)$^\spadesuit$ & Shallow & 36.1 & 35.4 & 36.7 & 12.5 & 54.4 & 46.7 & 8.3 & - & - & - & 88.2 & 60.5 & - & 14.7 \\
    \mname & DeepSeek-R1 & Text (Multi)$^\spadesuit$ & Shallow & 39.2 & 34.2 & 44.3 & 20.8 & 57.4 & 50.0 & 8.3 & - & - & - & 87.5 & 61.1 & - & 14.5 \\

    \mname & Qwen3-32B & Text (Multi)$^\spadesuit$ & Deep & 39.2 & 41.8 & 36.7 & 20.8 & 64.7 & 26.7 & \textbf{\textcolor{purple}{13.9}} & 84.2 & 75.9 & 76.0 & 80.8 & 59.7 & 35.1 & 14.2 \\
    \mname & Qwen3-235B & Text (Multi)$^\spadesuit$ & Deep & 45.6 & 43.0 & \textbf{\textcolor{purple}{48.1}} & \underline{41.7} & 63.2 & 50.0 & 11.1 & 88.0 & 77.0 & 78.4 & 82.9 & 60.6 & 37.7 & 14.8 \\
    \mname & DeepSeek-R1 & Text (Multi)$^\spadesuit$ & Deep & 45.6 &  44.3 & 46.8 & 33.3 & 67.7 & 46.7 & 11.1 & \underline{89.0} & \textbf{\textcolor{purple}{79.1}} & \textbf{\textcolor{purple}{79.8}} & 82.6 & 59.2 & 38.7 & 12.9 \\

    \mname & Qwen3-32B & Hybrid (Jina) & Deep   & 42.4 & 43.0 & 41.8 & 20.8 & \textbf{\textcolor{purple}{70.6}} & 33.3 & 11.1 & 87.0 & 78.1 & 78.8 & 83.2 & 59.4 & \underline{43.1} & 17.2 \\
    \quad - w/o planner & Qwen3-32B & Hybrid (Jina) & Deep  & 36.7 & 39.2 & 35.4 & 29.2 & 50.0 & 40.0 & 11.1 & - & - & - & 85.7 & 58.1 & 40.5 & 14.5\\
    \mname & Qwen3-235B & Hybrid (Jina) & Deep  & \underline{47.5} & \underline{46.8} & \textbf{\textcolor{purple}{48.1}} & \underline{41.7} & 64.7 & \underline{53.3} & \textbf{\textcolor{purple}{13.9}} & 86.7 & \textbf{\textcolor{purple}{79.1}} & \underline{79.4} & 82.2 & 61.2 & \textbf{\textcolor{purple}{45.7}} & 15.8 \\
    \quad - w/o planner & Qwen3-235B & Hybrid (Jina) & Deep  & 39.2 & 35.4 & 43.0 & 26.7 & 57.4 & 43.3 & \textbf{\textcolor{purple}{13.9}} & - & - & - & 88.0 & 58.2 & 41.2 & 13.9 \\
    \mname & DeepSeek-R1 & Hybrid (Jina) & Deep & \textbf{\textcolor{purple}{50.6}} & \textbf{\textcolor{purple}{53.2}} & \textbf{\textcolor{purple}{48.1}} & \textbf{\textcolor{purple}{45.8}} & \textbf{\textcolor{purple}{70.6}} & \textbf{\textcolor{purple}{56.7}} & 11.1 & \textbf{\textcolor{purple}{89.5}} & 77.4 & 78.6 & 82.9 & 58.0 & 41.0 & 14.1 \\
    \quad - w/o planner & DeepSeek-R1 & Hybrid (Jina) & Deep & 42.9 & 44.0 & 41.8 & 31.2 & 61.7 & 43.3 & 8.3 & - & - & - & 84.2 & 55.3 & 38.8 & 13.6 \\

    \bottomrule
\end{tabular}}
\caption{Main results on \dname. The best score is in \textcolor{purple}{boldface and colored} and second best is \underline{underlined}.  $^\heartsuit$ means BGE-M3 that for text retrieval and ColQwen for page retrieval.  $^\spadesuit$ refers to using Qwen3-embed, BGE-M3, and E5 for text retrieval. $^\clubsuit$ indicates using InternVL3.5-38B to extract relevant information from raw screenshots, as LLM backbones cannot read images.}
\vspace{-1em}
\label{tab:overall_main}
\end{table*}

\subsection{System Setting and Baselines}

\textit{\textbf{\quad Retrieval System}.} 
We evaluate 10 retrievers to implement our multimodal retrieval architecture ($\S$\ref{ssec:retrieval_arch}): \textbf{5 Text retrievers}: BM25 \citep{bm25}, E5~\citep{wang2022-e5}, BGE-M3~\citep{chen2024bgem3}, Qwen3-embedding-0.6B~\citep{zhang2025qwen3embedding}, and Qwen3-reranker-0.6B~\citep{zhang2025qwen3embedding}.
\textbf{5 Vision retrievers}: DSE~\citep{ma2024dse}, ColPali~\cite{faysse2024colpali}, ColQwen~\cite{faysse2024colpali}, and Jina-embedding-v4~\citep{günther2025jinaembeddingsv4} (multi-vector and dense variants).
We configure these retrievers in three paradigms:
\textbf{(1) Text-only}: Text retrievers encode both textual and visual chunks, with visual elements represented through VLM descriptions.
\textbf{(2) Vision-only}: Vision retrievers directly retrieve page screenshots without parsing.
\textbf{(3) Hybrid}: Text retrievers encode textual chunks while vision retrievers encode visual chunks, leveraging modality-specific strengths.
For multi-query (decomposing original question into multiple subqueries) aggregation, we employ Qwen3-reranker-0.6B for text retrieval reranking. For vision and hybrid retrieval, we accumulate equal passages per sub-query due to the absence of robust multimodal rerankers.

\textit{\textbf{\mname Configurations}.}
\mname is implemented with three parsing levels ($\S$\ref{ssec:deep_parsing}), all using the same backbone LLM (Qwen3-32B~\citep{yang2025qwen3}, Qwen3-235B~\citep{yang2025qwen3}, DeepSeek-R1~\citep{deepseekai2025deepseekr1}) but differing in document representation:
\textbf{(1) Parsing-free}: Direct page screenshot processing, requiring additional VLMs (InternVL3.5-38B~\citep{wang2025internvl35}) for information extraction from retrieved images.
\textbf{(2) Shallow Parsing}: OCR-extracted text split into fixed-length chunks without structural awareness.
\textbf{(3) Deep Parsing}: Layout-aware chunks with multimodal elements transcribed via VLM, preserving document structure and semantics.

\textit{\textbf{Baseline Systems}.}
We compare \mname against non-RAG and state-of-the-art document RAG systems:
\textbf{(1) Direct}: LLM responses without document context, testing parametric knowledge.
\textbf{(2) Long-context}: Full documents provided as context (truncated at 96k tokens), testing long-context understanding based on both shallow and deep parsed documents.
\textbf{(3) MDocAgent}~\citep{han2025mdocagent}: Multi-agent pipeline processing top-5 text text chunks (BGE-M3) and top-5 page screenshots (ColQwen).
\textbf{(4) M3DocRAG}~\citep{cho2024m3docrag}: Multimodal input combining query with top-10 ColPali-retrieved page screenshots.
\textbf{(5) ColQwen-gen}~\citep{faysse2024colpali}: Identical to M3DocRAG but using ColQwen retriever.

\subsection{Multimodal Retrieval System Results}

Table~\ref{tab:retrieval_main} presents retrieval performance across different paradigms on \dname and MMDocIR, evaluating both original questions and decomposed sub-queries. Fine-grained results are in Figure~\ref{fig:radar_chart_retrieval}.

\textit{\textbf{Retrieval Paradigm Comparison}.} 
Text-only retrieval shows hierarchical performance: sparse retrieval (BM25) performs worst, while dense retrievers (BGE-M3, Qwen3-Embedding) achieve substantial improvements. Vision-only retrieval, Jina-embedding-v4, even its dense variant, outperform other multi-vector models. Note that vision-only retrieval excels text-only method on MMDocIR's visual-centric tasks but underperforms on \dname's complex multi-document scenarios. Critically, hybrid retrieval consistently outperforms single-modality approaches. For instance, combining Qwen3-Embedding with Jina-embedding-v4 improves page recall by 8-12\% over either method alone, demonstrating that heterogeneous retrieval leverages complementary modality strengths.

\textit{\textbf{Sub-query Decomposition Impact}.} 
Query decomposition substantially enhances retrieval across all paradigms. Qwen3-Embedding's page-level recall improves from 42.1\% to 49.9\% (k=15) with sub-queries, while hybrid methods show similar gains. This confirms that iterative sub-query refinement can effectively chains evidence across documents and modalities, particularly crucial for multi-hop reasoning tasks.

\textit{\textbf{Granularity-Performance Tradeoff}.}
Performance degrades significantly with finer granularity: document-level recall exceeds 70\% for hybrid methods, but page- and layout-level recall drop by 30-40\%. This indicates that coarse-grained tasks are relatively easy to satisfy, whereas fine-grained retrieval task requires more adaptive and multi-granular retrieval strategies.

\subsection{Main Results: Multimodal Deep Research}

Table~\ref{tab:overall_main} validates our core contributions via systematic comparisons.

\textit{\textbf{Deep Research can largely improve multimodal understanding}.}
Direct answering achieves only 5-10\%; long-context processing ranges within 9-31\% despite document content of 96k tokens is given.  Results show that neither parametric knowledge nor brute-force context extension can handle research-level complexity. In comparison, \mname with multi-agent framework (with strategic planning, iterative search-refine loops, and progressive evidence accumulation) achieves score of 50.6\%.
The 20-40\% gap over long-context proves that deep research needs fundamentally multi-agent workflows, not just more context.

\textit{\textbf{Deep Parsing Outperforms Simplistic Parsing}.}
Relying on shallow parsing, \mname achieves 34-39\% accuracy, which is worse than using parsing-free screenshot processing (36-43\%). This confirms that simplistic OCR destroys visual semantics. \mname equipped with deep parsing and hybrid retrieval, which preserves layout and multimodal characteristics, reaches 42-50\%. The 10\% absolute gain validates that modality-specific preservation via deep parsing is essential for document understanding.

\textit{\textbf{Iterative Workflows Enable Deep Research}.}
Baseline RAG systems fail due to single-round and rigid retrieval: MDocAgent (15.8\%) cannot adaptively select strategies despite using dual modalities; M3DocRAG (7.0\%) and ColQwen-gen (5.7\%) are limited to page-level vision retrieval. In contrast, \mname achieves 50.6\% which is 3.4× better than MDocAgent, as it allows for the dynamic decomposition and refinement of queries over multiple steps, enabling the multi-hop reasoning and evidence synthesis.

\begin{figure}[t]
    \centering
    \includegraphics[width=1.0\linewidth]{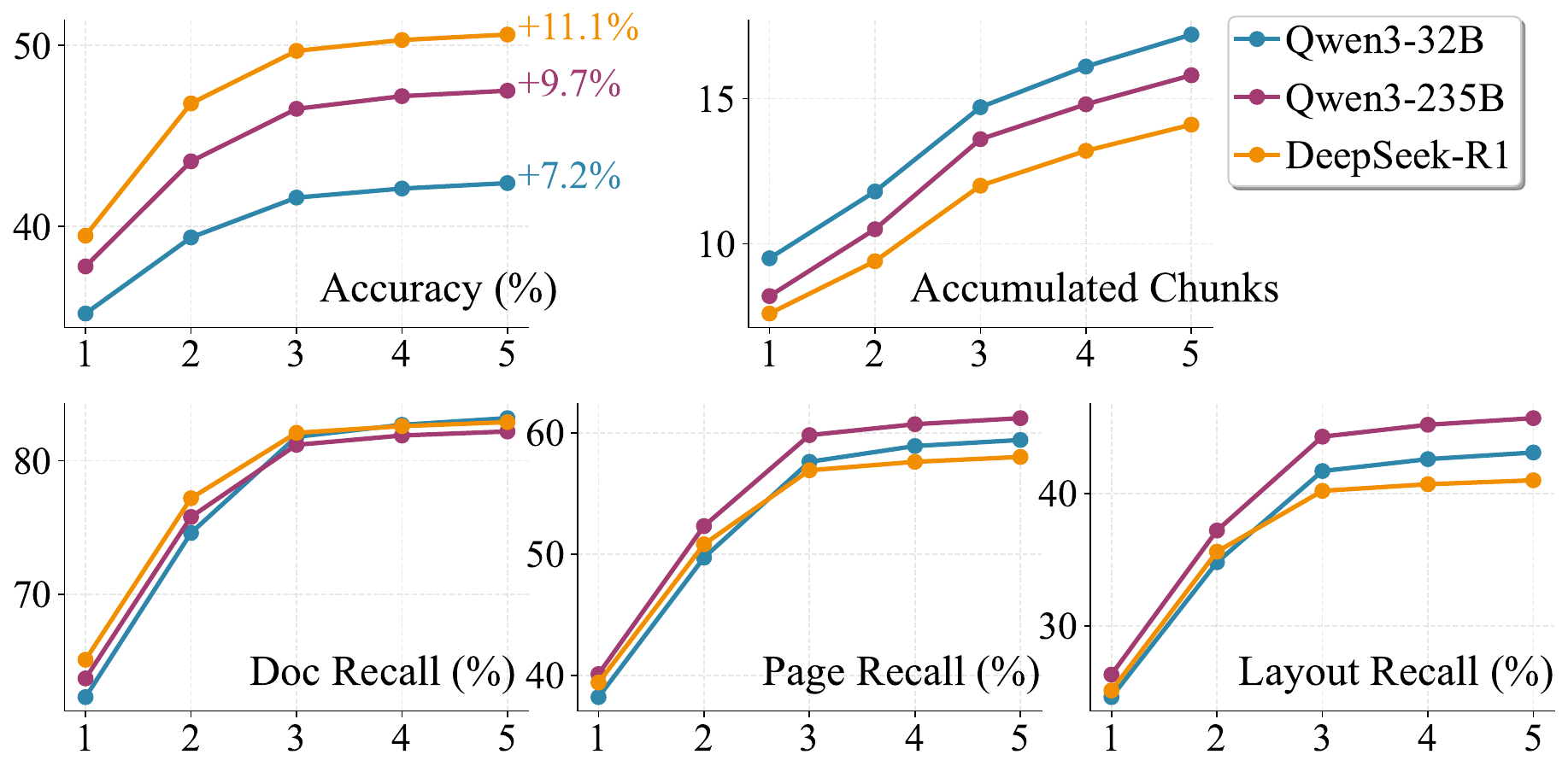}
    \vspace{-2.5em}
    \caption{Performance with increasing search depth.}
    \vspace{-2.0em}
    \label{fig:iterative_search}
\end{figure}

\subsection{Ablation Study}
We ablate components of \mname, and validate different architectural design as follows:

\textit{\textbf{Hybrid Retrieval Necessity}.}
Within our deep parsing framework, hybrid retrieval (42.4-50.6\%) outperforms text-only (39.2-45.6\%) by 3-5\%. This shows that even with high-quality textual transcriptions in both coarse and fine-grained level, some visual semantics is lost. In comparison, visual encoders that directly encode using chunk image, can capture irreplaceable semantics. This validates our hybrid retrieval design over single-modality approaches.

\textit{\textbf{Adaptive Planning Impact}.}
Removing the Planner causes the accuracy drop of 6-8\%. Without it, the system defaults to uniform chunk retrieval across given noisy document sets. In this setting, the system loses capabilities as follows: (1) strategic document filtering (critical for \dname's 12.7-document average for each question), and (2) query-adaptive granularity selection. This confirms that multi-granular representations require intelligent selection to be effective.

\textit{\textbf{Effect of Iterative Search}.}
Figure~\ref{fig:iterative_search} demonstrates how iterative refinement progressively improves retrieval and answer quality. Document-level recall jumps from 62-65\% (turn 1) to 75-82\% (turn 3), then plateaus at approximately 82-83\% by turn 5, indicating that the search-refine loop effectively identifies relevant documents within 2-3 iterations. Accuracy exhibits substantial gains: Qwen3-32B improves by 7.2\%, Qwen3-235B by 9.7\%, and DeepSeek-R1 by 11.1\% from turn 1 to turn 5. However, the steepest improvements occur between turns 1-3, after which gains diminish despite accumulating additional chunks (from 7-10 at turn 1 to 14-17 at turn 5), suggesting increasing redundancy. Notably, stronger models achieve higher accuracy with fewer chunks (DeepSeek-R1: 14.1 vs Qwen3-32B: 17.2 at turn 5), indicating more efficient evidence selection. Fine-grained layout recall follows similar patterns, improving from 24-26\% to 40-45\%. These results validate our iterative search design while suggesting that 3 iterations offer an optimal balance between retrieval quality and computational efficiency.

\begin{table}[t]
\renewcommand{\arraystretch}{0.9}
\setlength{\tabcolsep}{2.5pt}
\centering
\resizebox{1.0\linewidth}{!}{
\begin{tabular}{l@{\hskip -1.0pt}c@{\hskip 1.0pt}|c@{\hskip 2.0pt}c@{\hskip 2.0pt}c@{\hskip 2.0pt}c@{\hskip 0.0pt}|c@{\hskip 0.0pt}c@{\hskip 2.0pt}cc}
    \toprule
    \multicolumn{2}{c|}{Parsing Setting } & \multicolumn{4}{c|}{Parsing Latency and Cost} & \multicolumn{4}{c}{\mname Latency} \\

    \cmidrule{1-2} \cmidrule{3-6} \cmidrule{7-10}
    
    Level & Type & Parse & Index & Caption & Store & Plan & Search & Refine & Answer \\
    \midrule
    \multirow{2}{*}{free} & vision:dense & - & 08:56 & - & 0.15 & - & 05:34 & 19:50:16 & 03:30:42 \\

    & vision:multi & - & 39:14 & - & 4.33 & - & 59:47 & 20:32:23 & 03:41:12 \\
    \midrule
    shal' & text:dense & 18:47 & 09:50 & - & 0.37 & - & 10:31 & 03:29:37 & 01:57:32 \\
    \midrule
    \multirow{2}{*}{deep} & text:dense & 01:21:21 & 12:47 & 01:12:32 & 0.34 & 01:48:32 & 10:54 & 04:14:30 & 02:27:28 \\
    & hybrid:dense & 01:21:21 & 14:40 & 01:12:32 & 0.43 & 02:01:16 & 02:19 & 04:23:20 & 02:09:42 \\
    \bottomrule
\end{tabular}}
\caption{Efficiency analysis on different parsing setting and their effect on \mname systems. The latency is measured by HH:MM:SS and storage is in GB.}
\vspace{-1em}
\label{tab:efficiency_stats}
\end{table}

\subsection{Parsing and Deep Research Efficiency}

Table~\ref{tab:efficiency_stats} analyzes computational tradeoffs across parsing strategies on \dname's 304 documents (6,177 pages, 4,146 figures, and 2,739 tables) and deep research latency on 158 questions.

\textit{\textbf{Parsing-time vs. Research-time Trade-off}.} Deep parsing requires around 2.5h upfront: much slower than parsing-free (9-39m), which primarily due to MinerU's layout analysis and VLM-based visual element transcription. 
However, this investment dramatically reduces research-time latency. Parsing-free methods spend 80\% of inference time (20h) extracting information from raw screenshots, while deep parsing completes the same tasks in 4h. This 5× speedup during research validates that one-time multimodal transcription is more efficient than repeated visual processing.

\textit{\textbf{Storage and Embedding Considerations}.}
Multi-vector embeddings consume 10-20× more storage and 4x more indexing time than dense variants, reflecting the tradeoff between retrieval quality and resource requirements. Note that multi-vector embedding storage and similarity computation (maxsim in late interaction mechanism) is not supported in current vector database, thus causing much longer indexing and searching latency.

\textit{\textbf{Research Phase Breakdown}.}
For deep research workflows, latency distributes across planning (2h), iterative search-refine loops (4h), and answer generation (2h). The search-refine phase dominates due to multiple retrieval rounds and evidence accumulation.

\section{Conclusion}
\label{sec:conclusion}

We introduced Doc-Researcher, addressing the critical limitation that existing deep research systems cannot process multimodal documents. Our work identifies and resolves three fundamental gaps in current systems.
First, we developed a deep multimodal parsing framework that preserves layout structure and modality-specific characteristics, achieving 11.4\% improvement over simplistic OCR approaches. Second, we designed a systematic retrieval architecture supporting multiple paradigms (text/vision/hybrid) and granularities (chunk/page/full/summary). Third, we implement iterative multi-agent workflows that mirror human research processes through planning, search-refine loops, and progressive synthesis, leading to 3.4× better performance than existing baselines. Beyond the system, we contributed M4DocBench, the first benchmark evaluating multimodal, multi-hop, multi-document, and multi-turn research capabilities with 158 expert-annotated questions spanning 4 domains.
In the future, we will explore three directions: developing compact multimodal embeddings to reduce computational overhead, improving cross-modal reasoning mechanisms for better evidence synthesis, and extending to active learning that adapts retrieval strategies based on user interaction patterns. 
These advances would further enhance efficiency/effectiveness of multimodal document research.


\bibliographystyle{ACM-Reference-Format}
\bibliography{sample-base-backup}

\appendix

\section{\dname: Supplementary Materials}
\label{appendix:dataset}

\quad \textit{\textbf{Statistical Details}}.
In this section, we provide more fine-grained statistics of \dname, which supplements analysis in Section~\ref{ssec:benchmark_stats}.
In Table~\ref{tab:doc_stats}, we report the statistical details about multimodal documents in \dname. Note that documents of all domains contain considerable figures and tables.
In Table~\ref{tab:dataset_main}, we provide more fine-grained statistics of \dname by domains as well as statistics of question, document, evidence, answer, etc.

\begin{figure*}[t]
    \centering
    
    \begin{subfigure}{\linewidth}
        \centering
        \caption{The question requires integrating textual and tabular evidence from multiple market analysis reports to compare fund-raising and exit-return proportions across different months. The reasoning involves multi-hop retrieval and numerical computation across documents. }
        \vspace{-0.2em}
        \includegraphics[width=0.99\linewidth]{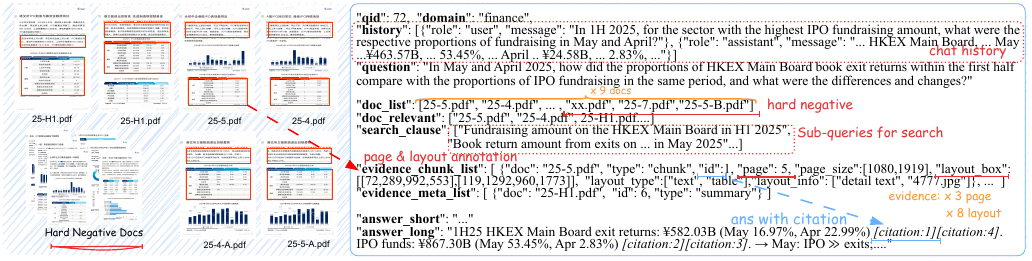}
        \label{fig:a}
    \end{subfigure}
    
    
    \begin{subfigure}{\linewidth}
        \centering
        \caption{This example illustrates a cross-document, cross-modal, multi-hop reasoning task. The model needs to extract and compare structured program information across multiple brochures to determine the shortest graduation time under the given limitation. }
        \vspace{-0.2em}
        \includegraphics[width=0.99\linewidth]{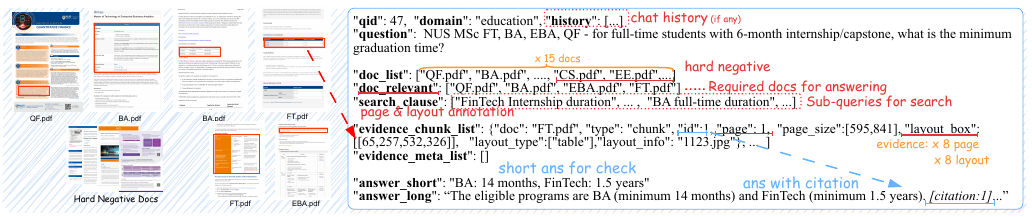}
        \label{fig:b}
    \end{subfigure}
    \vspace{-1.5em}
    \caption{Demonstration of two typical annotations of \dname. Note that they are simplified for brevity.}
    \label{fig:annotations}
\end{figure*}

\textit{\textbf{Annotation Demonstration}}.
Although Figure~\ref{fig:task_view} shows a typical annotation of \dname, we provide two more annotation examples as shown in Figure~\ref{fig:annotations}. This is to facilitate readers' understanding on \dname and its annotation process in Section~\ref{ssec:annotation_process}. 

\begin{table}[ht]
\centering
\small
\renewcommand{\arraystretch}{0.8}
\begin{tabular}{l|rrrrrrr}
    \toprule
    \multirow{2}{*}{Domain} & Doc & Avg & Avg & Avg & Avg & Avg \\
    & Count & Page & Word & Fig. & Tab. & Eq. \\
    \midrule
    Research   & 166 & 26.7 & 21.8k & 17.3 & 11.5 & 2.1 \\
    Insurance  & 43  & 11.0 & 5.5k  & 10.8 & 5.4  & 0.0 \\
    Education  & 56  & 7.5  & 3.9k  & 4.3  & 4.3  & 0.0 \\
    Finance    & 38  & 22.2 & 11.2k & 14.4 & 9.5  & 0.0 \\
    \midrule
    \textbf{Overall} & \textbf{303} & \textbf{20.4} & \textbf{14.8k} & \textbf{13.6} & \textbf{9.0} & \textbf{1.2} \\
    \bottomrule
\end{tabular}
\caption{Document statistics across different domains}
\vspace{-1.0em}
\label{tab:doc_stats}
\end{table}

\begin{table}[t]
\small
\renewcommand{\arraystretch}{0.8 }
\setlength{\tabcolsep}{2.5pt}
    \centering
    \resizebox{\linewidth}{!}{%
\begin{tabular}{l|l|ccl}
\toprule
& \textbf{Model} & \textbf{Size} & \textbf{Base Model} & \textbf{HuggingFace Checkpoint}                               \\ 
\midrule

\parbox[t]{2.5mm}{\multirow{5}{*}{\rotatebox[origin=c]{90}{Text}}} 

& BM25 & -  & - & - \\
& BGE-M3 & 0.56B & XLM-RoBERTa-large & \href{https://huggingface.co/BAAI/bge-m3}{BAAI/bge-m3} \\ 
& E5 & 0.34B & BERT-large & \href{https://huggingface.co/intfloat/e5-large-v2}{intfloat/e5-large-v2}  \\ 
& Qwen3-embed &  0.6B & Qwen3-0.6B-Base & \href{https://huggingface.co/Qwen/Qwen3-Embedding-0.6B}{Qwen/Qwen3-Embedding-0.6B}  \\
& Qwen3-reranker & 0.6B & Qwen3-0.6B-Base & \href{https://huggingface.co/Qwen/Qwen3-Reranker-0.6B}{Qwen/Qwen3-Reranker-0.6B} \\

\midrule

\parbox[t]{2.5mm}{\multirow{4}{*}{\rotatebox[origin=c]{90}{Visual}}} &
DSE$_{\mathrm{wiki-ss}}$ & 4.15B & Phi-3-Vision & \href{https://huggingface.co/Tevatron/dse-phi3-v1.0}{Tevatron/dse-phi3-v1.0} \\
& ColPali & 2.92B & PaliGemma & \href{https://huggingface.co/vidore/colpali}{vidore/colpali} \\
& ColQwen2 & 2.21B &  Qwen2-VL-2B-Instruct & \href{https://huggingface.co/vidore/colqwen2-v1.0}{vidore/colqwen2-v1.0} \\
& Jina-embed-v4 & 3.75B &  Qwen2.5-VL-3B-Instruct & \href{https://huggingface.co/jinaai/jina-embeddings-v4}{jinaai/jina-embeddings-v4} \\

\bottomrule
\end{tabular}
}
\caption{Checkpoints for Text and Vision Retrievers.}
\vspace{-1.0em}
\label{tab:retriever-implementation-details}
\end{table}

\begin{table*}[htbp]
\centering
\renewcommand{\arraystretch}{0.8}
    \resizebox{0.9\linewidth}{!}{%
    \begin{tabular}{l|m{2.6cm}m{2.8cm}m{2.75cm}m{2.75cm}|m{2.6cm}}
        \toprule
        \textbf{Metric} & \textbf{Education} & \textbf{Finance} & \textbf{Insurance} & \textbf{Research} & \textbf{Overall} \\
        \midrule

        \multicolumn{6}{l}{\textbf{Question Statistics} \textcolor{gray}{\textit{: at least one evidence, evidence of multiple modalities and from multi-documents, multi-turn conversations}}} \\
        
        \#Question & 30 & 36 & 24 & 68 & 158 \\

        TXT/TAB/I & 24 / 22 / 0 & 34 / 24 / 28 & 18 / 22 / 12 & 50 / 20 / 30 & 126 / 88 / 70 \\
        Multi-mod/doc/turn &  22 / 30 / 4 & 36 / 36 / 6 & 18 / 22 / 16 & 36 / 56 / 32 & 112 / 144 / 58  \\

        \midrule
        
        \multicolumn{6}{l}{\textbf{Document Statistics per Question} \textcolor{gray}{\textit{: min / max / median / \underline{avg}}}} \\
        \#Docs (total) & 4 / 16 / 8 / \underline{7.8} & 9 / 12 / 9 / \underline{10.0} & 9 / 42 / 11 / \underline{13.8} & 5 / 22 / 17 / \underline{15.9} & 4 / 42 / 11 / \underline{12.7} \\
        \#Docs (relevant) & 2 / 6 / 2 / \underline{3.0} & 3 / 7 / 3 / \underline{3.9} & 1 / 10 / 3 / \underline{4.6} & 1 / 10 / 4 / \underline{3.8} & 1 / 10 / 3 / \underline{3.8} \\
        \midrule
        
        \multicolumn{6}{l}{\textbf{Evidence Statistics per Question} \textcolor{gray}{\textit{: min / max / median / \underline{avg}}}} \\
        \#Total Page & 8/174/55/ \underline{72.5} & 151/357/207/ \underline{222.6} & 123/461/165/ \underline{195.8} & 132/836/484/ \underline{493.0} & 8/836/228/ \underline{306.4} \\
        
        \#Relevant Page & 2 / 8 / 3 / \underline{3.3} & 3 / 19 / 6 / \underline{6.7} & 2 / 31 / 4 / \underline{9.6} & 1 / 19 / 7 / \underline{7.9} & 1 / 31 / 5 / \underline{7.0} \\
        
        \#Layout & 2 / 8 / 5 / \underline{4.9} & 6 / 48 / 12 / \underline{14.8} & 2 / 40 / 6 / \underline{11.7} & 1 / 157 / 16 / \underline{20.6} & 1 / 157 / 8 / \underline{14.8} \\

        \midrule
        \multicolumn{6}{l}{\textbf{Evidence Modality Distribution} \textcolor{gray}{\textit{: text (body text, title, equations, etc) / table / image (figure, chart, infographic, etc) }}} \\
        \% by \#Chunk & 52.4 / 47.6 / 0.0 & 44.0 / 38.0 / 18.0 & 49.6 / 42.6 / 7.8 &  83.8 / 6.8 / 9.4 &  67.5 / 16.6 / 15.9  \\
        \% by Bbox Area  & 28.2 / 71.8 / 0.0 & 41.5 / 22.1 / 36.4 & 9.2 / 83.8 / 7.0 & 69.0 / 18.4 / 12.6 & 40.9 / 30.2 / 28.9 \\ 
        
        \midrule
        
        \multicolumn{6}{l}{\textbf{Answer Statistics: length (words)} \textcolor{gray}{\textit{: min / max / median / \underline{avg}}}} \\
        Short Answer & 13 / 303 / 26 / \underline{49.1} & 35 / 383 / 85 / \underline{108.7} & 13 / 80 / 50 / \underline{48.4} & 5 / 117 / 32 / \underline{36.1} & 5 / 383 / 44 / \underline{59.7} \\
        
        Long answer & 40/747/122/ \underline{174.7} & 185/698/478/ \underline{464.8} & 115/650/266/ \underline{314.2} & 23/550/112/ \underline{183.4} & 23/747/223/ \underline{272.0} \\
        \midrule
        
        \multicolumn{6}{l}{\textbf{Other Statistics} \textcolor{gray}{\textit{: min / max / median / \underline{avg}}}} \\

        \#Multi-turn & 2 / 2 / 2 / \underline{2.0} & 2 / 2 / 2 / \underline{2.0} & 2 / 4 / 3 / \underline{2.6} & 2 / 3 / 2 / \underline{2.2} & 2 / 4 / 2 / \underline{2.3} \\
        
        \#Search clauses & 2 / 8 / 3 / \underline{3.7} & 2 / 12 / 5 / \underline{5.2} & 3 / 10 / 5 / \underline{4.8} & 1 / 6 / 3 / \underline{3.1} & 1 / 12 / 3 / \underline{3.9} \\
        
        \bottomrule
    \end{tabular}}
    \caption{Statistical details of \dname by domains and }
    \label{tab:dataset_main}
\end{table*}

\begin{figure*}[t]
    \centering
    \includegraphics[width=0.99\linewidth]{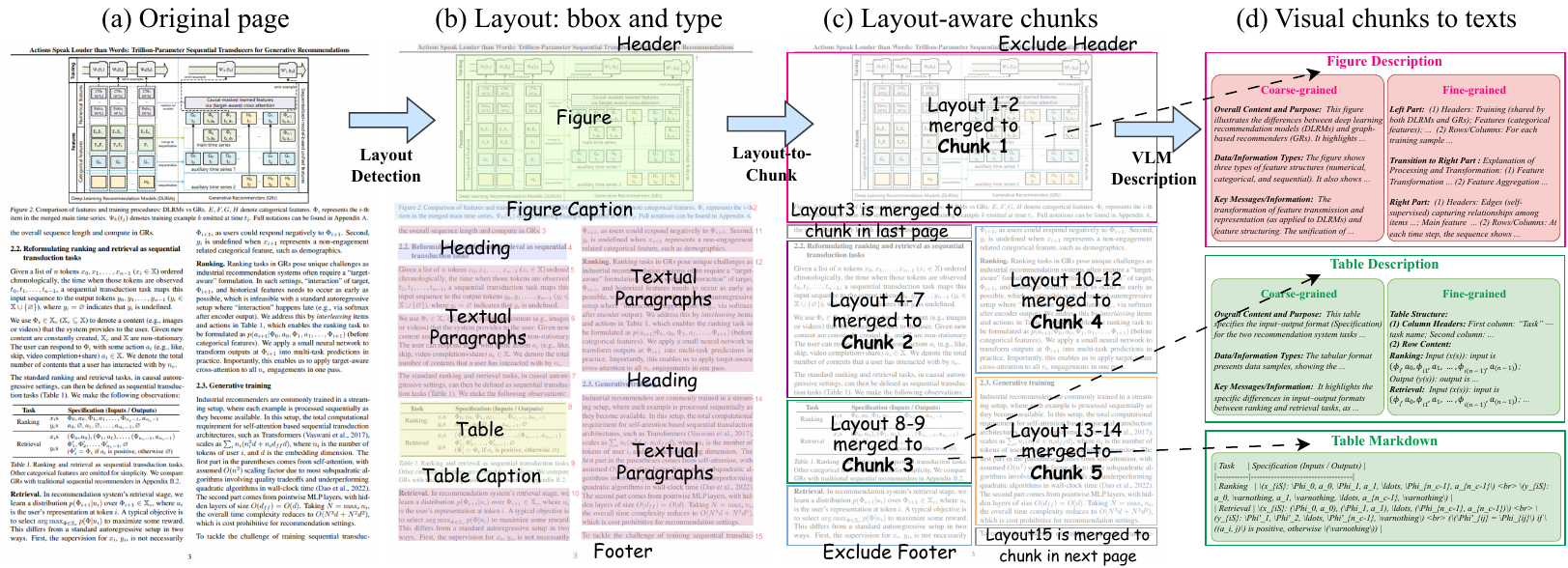}
    \vspace{-0.5em}
    \caption{The overview of layout-aware chunking process and multimodal-to-text transcription.}
    \label{fig:layout_chunk}
\end{figure*}

\begin{figure*}[t]
    \centering
    \includegraphics[width=0.99\linewidth]{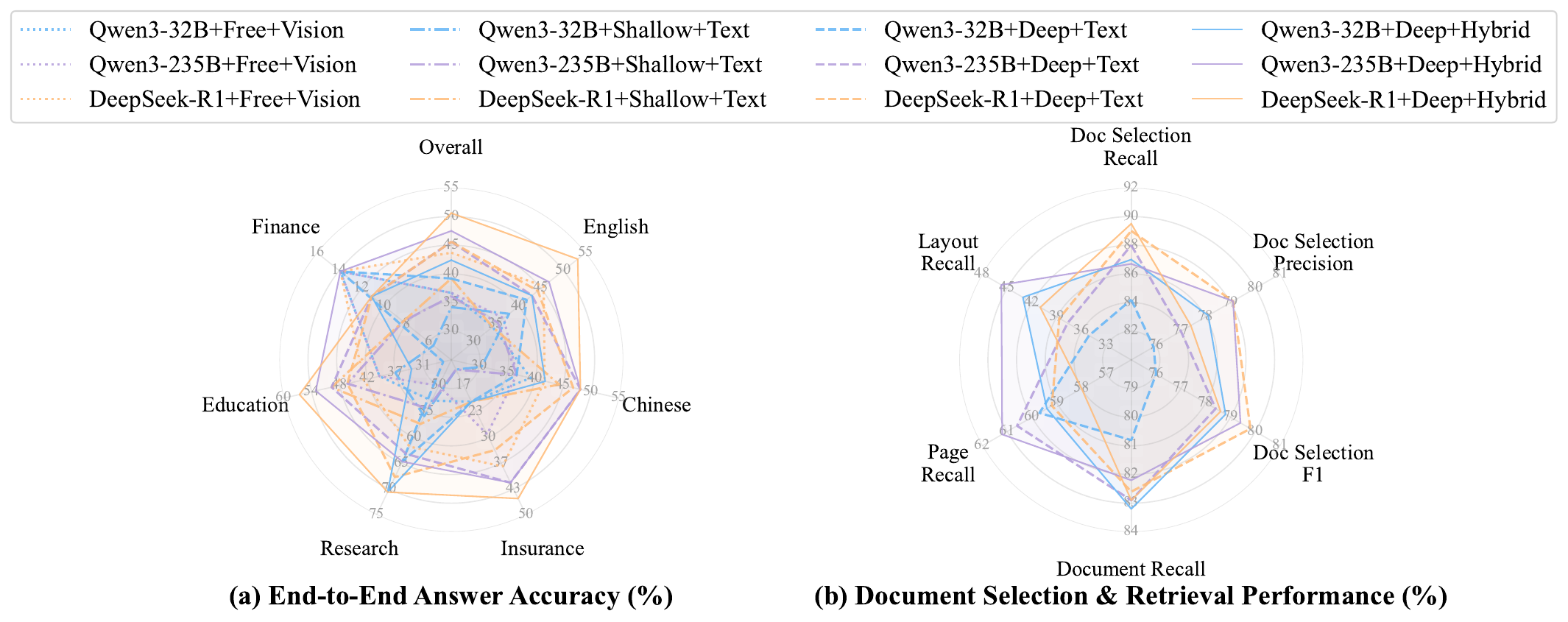}
    \vspace{-0.5em}
    \caption{Accuracy and retrieval performance of \mname using different backbone LLMs and retrieval schemes.}
    \label{fig:radar_chart_e2e}
\end{figure*}

\begin{figure*}[t]
    \centering
    
    \includegraphics[width=0.99\linewidth]{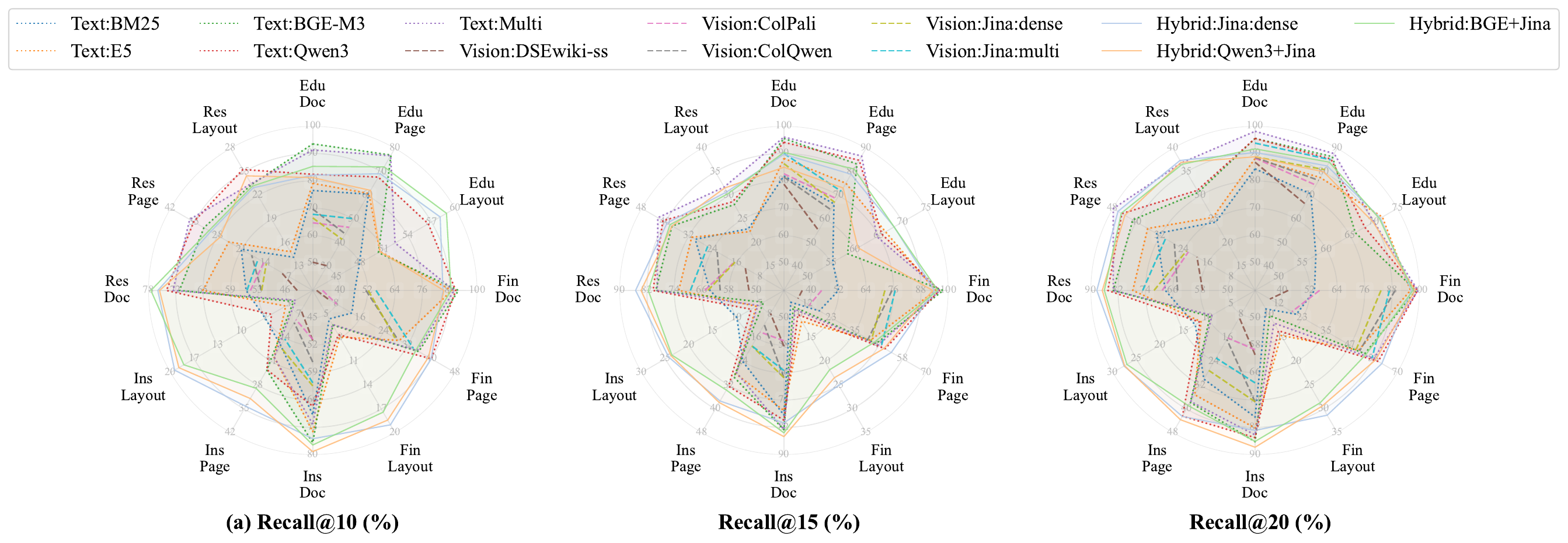}
    
    \vspace{0.5em}
    
    \includegraphics[width=0.9\linewidth, trim=0 1cm 0 0, clip]{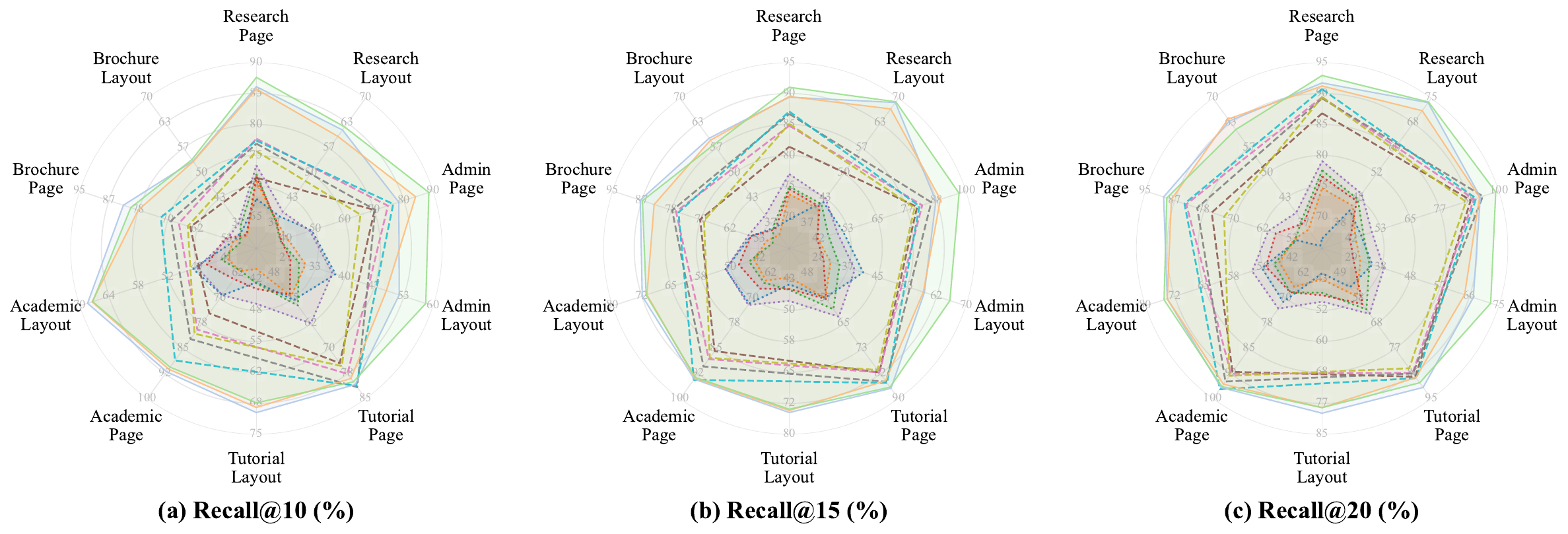}

    \hspace{2mm} 
    \includegraphics[width=0.99\linewidth]{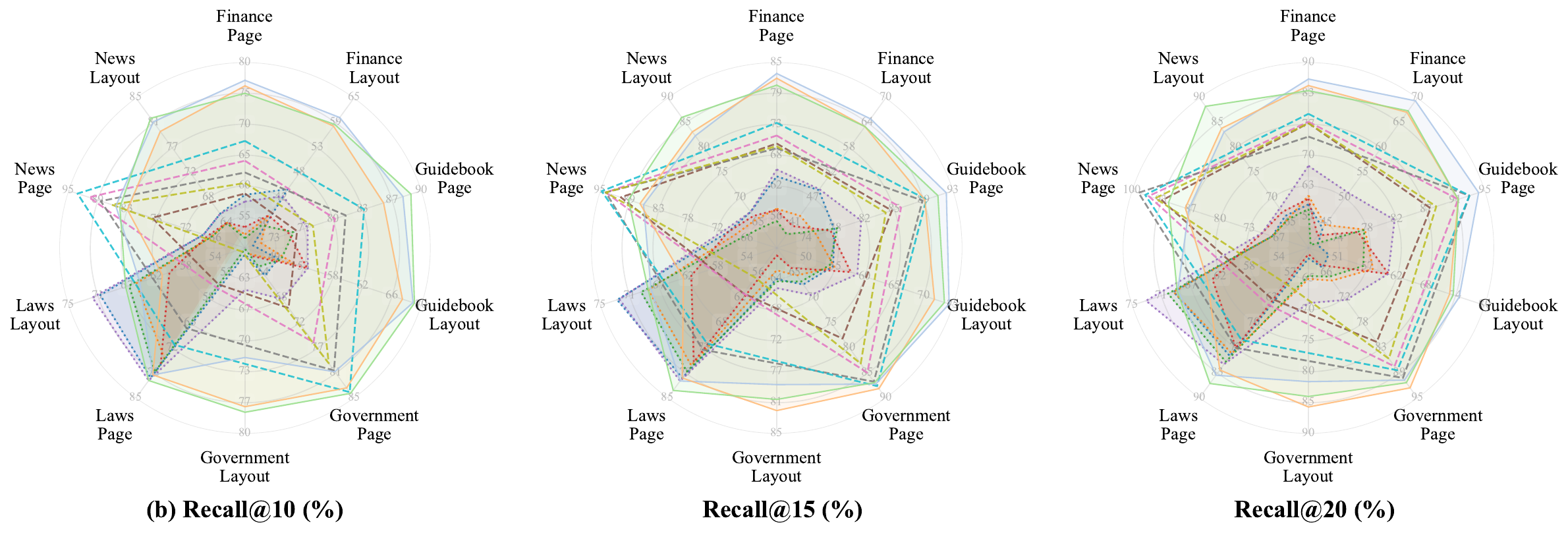}
    \vspace{-0.5em}
    \caption{(a) Retrieval performance (break-down by 4 domains) using text/vision/hybrid scheme on \dname. (b) Retrieval performance (break-down by 10 domains) using text/vision/hybrid scheme on MMDocIR. Note that recall of layout is not applicable to vision-only scheme, as the retrieval is at page level. MMDocIR focuses on single-document VQA, hence document recall is not applicable.}
    \label{fig:radar_chart_retrieval}
\end{figure*}

\section{Implementation Details}
\label{appendix:impl_details}

\textbf{\textit{Multimodal Document Parsing}}. We utilize MinerU v2.0.6 for document parsing. The chunking strategy segments text when: (1) length exceeds 300 tokens, (2) page boundaries are crossed, or (3) section headers are encountered. Images and tables are extracted as independent units. \textbf{Image Processing}: Visual content was processed using \href{https://huggingface.co/Qwen/Qwen2.5-VL-7B-Instruct}{Qwen2.5-VL-7B-Instruct}\footnote{Qwen2.5-VL-7B-Instruct is served using \href{https://github.com/vllm-project/vllm}{vLLM} for efficient and stable inference.} for description generation. Visual chunks of image size less than 10KB are excluded. This is to filter decorative icons and insignificant graphics, as well as relieving computational overhead. \textbf{OCR Processing}: PDF pages were converted using PyMuPDF (fitz) and subjected to Tesseract OCR with a chunk size of 140 characters for structured text extraction.

\textbf{\textit{Multimodal Indexing and Data Persistence}}. 
We deploy \textit{MySQL database} to store document metadata, including document summaries, structured full-text, chunks (both text and cropped images), pages (both text and screenshots), and layout bounding boxes (bbox coordinates). 
We summarize the checkpoints for retriever deployments in Table~\ref{tab:retriever-implementation-details}. Retrievers are used to convert passages of different modalities and granularities into vector representations.
We deploy \textit{Milvus vector database} to maintain dense vectors for textual and visual chunks, and page screenshots. Document names serve as filtering conditions for retrieval operations. For \textit{multi-vector storage}, we save them locally as np.savez\_compressed, with each document stored as an individual NPZ file for efficient retrieval.

\textbf{\textit{Deep Research Framework}}.
We implement \mname, the multi-agent deep research workflow using LangGraph\footnote{\url{https://www.langchain.com/langgraph}}.

\section{Extended Experimental Results}
\label{appendix:exp_results}

Figure~\ref{fig:radar_chart_e2e} depicts the accuracy and performance of \mname according to different domains and granularities.
Figure~\ref{fig:radar_chart_retrieval}a depicts the retrieval performance on \dname by 4 different domains and 3 different top-$k$ settings.
Figure~\ref{fig:radar_chart_retrieval}b depicts the retrieval performance on MMDocIR by 10 different domains and 3 different top-$k$ settings. This is extended retrieval results of Table~\ref{tab:retrieval_main}.

\end{document}